\documentclass{aa}  

\usepackage{xcolor}
\usepackage{graphicx}
\usepackage{comment}
\usepackage{txfonts}

\newcommand{\carmenes}{CARMENES}
\newcommand{\caracal}{CARACAL}
\newcommand{\gj}{GJ 1002}
\newcommand{\tyc}{TYC 8998 b}
\newcommand{\toto}{ExoMol \textsc{Toto}}
\newcommand{\ptr}{\textit{petitRADTRANS}}
\newcommand{\emcee}{\textsc{emcee}}

\newcommand{\isotope}[2]{\textsuperscript{#2}#1}
\newcommand{\lvmr}[1]{\ensuremath{\log_{10}{\left(\mathrm{VMR}_{#1}\right)}}}

\newcommand{\kms}{\ensuremath{\mathrm{km\ s}^{-1}}}
\newcommand{\K}{\ensuremath{\mathrm{K}}}
\newcommand{\sig}{\ensuremath{\sigma{}}}

\newcommand{\rj}{\ensuremath{R_\mathrm{J}}}
\newcommand{\rs}{\ensuremath{R_{\odot}}}

\newcommand{\tp}{$T$-$P$}
\newcommand{\teff}{\ensuremath{T_\mathrm{eff}}}
\newcommand{\logg}{\ensuremath{\log{g}}}
\newcommand{\metal}{\ensuremath{[\mathrm{M}/\mathrm{H}]}}
\newcommand{\turb}{\ensuremath{v_\mathrm{turb}}}
\newcommand{\lnlike}{\ensuremath{\ln{\mathcal{L}}}}

\newcommand{\snr}{\ensuremath{\mathrm{S/N}}}
\newcommand{\tint}{\ensuremath{t}}
\newcommand{\snrP}{\ensuremath{\left( \mathrm{S/N} \right)_{\mathrm{pix},\mathrm{planet}}}}
\newcommand{\snrS}{\ensuremath{\left( \mathrm{S/N} \right)_{\mathrm{pix},\mathrm{star}}}}
\newcommand{\contrast}{\ensuremath{c_{\lambda_0}}}
\newcommand{\fluxSE}{\ensuremath{F_{\lambda_0,\mathrm{star}}^{\bigoplus}}}
\newcommand{\area}{\ensuremath{A}}
\newcommand{\dlpix}{\ensuremath{\Delta{}\lambda{}_{\mathrm{pix}}}}
\newcommand{\ephot}{\ensuremath{E_{\lambda_0}}}
\newcommand{\nphotS}{\ensuremath{N_{\mathrm{pix},\mathrm{star}}}}
\newcommand{\npix}{\ensuremath{n_{\mathrm{pix}}}}

\begin{document} 

\title{ Measuring titanium isotope ratios in exoplanet atmospheres
    }

\titlerunning{Measuring titanium isotope ratios in exoplanet atmospheres}

\author{    Dilovan B. Serindag\inst{1},
            Ignas A. G. Snellen\inst{1},
            \and
            Paul Molli\`{e}re\inst{2}
        }

\institute{ Leiden Observatory, Leiden University, Postbus 9513, 2300 RA Leiden, The Netherlands
            \and
            Max-Planck-Institut f\"{u}r Astronomie, K\"{o}nigstuhl 17, 69117 Heidelberg, Germany
        }
        
\authorrunning{Serindag et al. (2021)}

\date{}
 
\abstract
{
    Measurements of relative isotope abundances can provide unique insights into the formation and evolution histories of celestial bodies, tracing various radiative, chemical, nuclear, and physical processes. In this regard, the five stable isotopes of titanium are particularly interesting. They are used to study the early history of the solar system, and their different nucleosynthetic origins help constrain Galactic chemical models. Additionally, titanium's minor isotopes are relatively abundant compared to those of other elements, making them more accessible for challenging observations, such as those of exoplanet atmospheres.
}
{
    We aim to assess the feasibility of performing titanium isotope measurements in exoplanet atmospheres. Specifically, we are interested in understanding whether processing techniques used for high-resolution spectroscopy, which remove continuum information of the planet spectrum, affect the derived isotope ratios. We also want to estimate the signal-to-noise requirements for future observations.
}
{
    We used an archival high-dispersion \carmenes{} spectrum of the M-dwarf \gj{} as a proxy for an exoplanet observed at very high signal-to-noise. Both a narrow (7045--7090 \AA{}) and wide (7045--7500 \AA{}) wavelength region were defined for which spectral retrievals were performed using \ptr{} models, resulting in isotope ratios and uncertainties. These retrievals were repeated on the spectrum with its continuum removed to mimic typical high-dispersion exoplanet observations. The \carmenes{} spectrum was subsequently degraded by adding varying levels of Gaussian noise to estimate the signal-to-noise requirements for future exoplanet atmospheric observations.
}
{
    The relative abundances of all minor Ti isotopes are found to be slightly enhanced compared to terrestrial values. Loss of continuum information from broadband filtering of the stellar spectrum has little effect on the isotope ratios. For the wide wavelength range, a spectrum with signal-to-noise of 5 is required to determine the isotope ratios with relative errors $\lesssim$10\%. Super Jupiters at large angular separations from their host star are the most accessible exoplanets, requiring about an hour of observing time on 8-meter-class telescopes, and less than a minute of observing time with the future Extremely Large Telescope.
}
{}

\keywords{  planets and satellites: gaseous planets --
            planets and satellites: atmospheres --
            planets and satellites: composition --
            stars: abundances --
            stars: individual (\gj{}) --
            techniques: spectroscopic
        }

\maketitle

\section{Introduction} \label{sec:intro}

The relative abundances of isotopes in a given environment are determined by various radiative, chemical, nuclear, and physical processes that occurred throughout its history. Understanding the effects of these processes on the isotope ratios can therefore trace the formation and evolution of astronomical objects. For instance, preferential Jeans escape of protium (\isotope{H}{1}) and bombardment by deuterium-rich comets are invoked to explain the enhanced deuterium-to-hydrogen (D/H) ratio\footnote{Throughout this paper, we exclusively refer to number abundance ratios. For two species A and B, we abbreviate this using A/B.} of Earth's ocean water compared to the protosolar nebula \citep[e.g.,][]{genda2008,hartogh2011}. Recently, work has begun to plan and perform isotope measurements in exoplanet atmospheres. For instance, \citet{lincowski2019} and \citet{morley2019} discuss the feasibility of determining hydrogen and oxygen isotope ratios in exoplanet atmospheres using the upcoming James Webb Space Telescope. \citet{molliere2019a} studied the efficacy of determining isotope ratios using ground-based high-resolution (${\mathcal{R} \sim 100\, 000}$) spectroscopy, and conclude that while D/H measurements will only be possible with the future Extremely Large Telescope, current 8-m-class telescopes should be capable of determining carbon isotope ratios in exoplanets. Indeed, \citet{zhang2021} measured an isotope ratio in an exoplanet for the first time, finding a super-terrestrial \isotope{C}{13}/\isotope{C}{12} value for the young super-Jupiter \tyc{} using medium-resolution (${\mathcal{R}\sim4500}$) integral field spectroscopy. Since this planet orbits well beyond the CO snowline, they suggest its enhancement in \isotope{C}{13} could be due to accretion of ices enriched in \isotope{C}{13} by isotope fractionation processes. This may mark the beginning of using isotope ratios to probe formation histories of planets beyond our solar system.

Titanium has five stable isotopes -- \isotope{Ti}{46}, \isotope{Ti}{47}, \isotope{Ti}{48}, \isotope{Ti}{49}, \isotope{Ti}{50} -- with telluric relative abundances of 8.25\%, 7.44\%, 73.72\%, 5.41\%, and 5.18\% \citep{meija2016}. With about 25\% of its atoms approximately equally partitioned among the minor (less-abundant) isotopes, Ti compares favorably to hydrogen (\citealt{wood2004}; \citealt{linsky2006}; \citealt{altwegg2015}, and references therein), carbon \citep[][]{milam2005,asplund2009,meija2016}, and oxygen (\citealt{ayres2013}, and references therein; \citealt{romano2017}, and references therein), whose minor isotopes have relative abundances $\lesssim$2\% in the solar system and local interstellar medium. The isotopes of titanium have different nucleosynthetic origins, with oxygen and silicon burning in massive stars thought to be the main source of \isotope{Ti}{46} and \isotope{Ti}{47}, while \isotope{Ti}{48}, \isotope{Ti}{49}, and \isotope{Ti}{50} are thought to be mostly produced in type Ia and/or type II supernovae \cite[][and references therein]{hughes2008}. It is therefore expected that stars that formed in different environments may exhibit variations in Ti isotope ratios of a factor $\sim$2 or more \citep{hughes2008}. Indeed, observational studies targeting TiO isotopologue\footnote{Molecules comprised of different atomic isotopes, for instance, \isotope{Ti}{47}\isotope{O}{16} versus \isotope{Ti}{48}\isotope{O}{16}.} features in K- and M-dwarfs have determined relative deviations of tens of percent in Ti isotope ratios compared to terrestrial values \citep{wyckoff1972,lambert1977,clegg1979,chavez2009,pavlenko2020}. Such measurements have, in turn, been used to constrain Galactic chemical models \citep[e.g.,][]{hughes2008}.

Interestingly, relative abundances of Ti isotopes are also used to study the early evolution of the solar system. For instance, variations in isotope ratios across different populations of meteorites and other solar system bodies are used to study (in)homogeneity and thermal processing in the solar protoplanetary disk \citep[e.g.,][]{leya2008,trinquier2009}. Conversely, the similarity in telluric and lunar Ti isotope ratios places constraints on Moon formation theories \citep{zhang2012}. We note, however, that the relative variations reported in the solar system are $\lesssim$0.1\% -- orders of magnitude smaller than those found in stellar populations.

Both close-orbiting and young gas giants can have sufficiently high temperatures (${T \gtrsim 1500}$--$2000$ K) to allow gaseous TiO to persist in their atmospheres \citep[e.g.,][]{hubeny2003,fortney2008,spiegel2009,gandhi2019}, possibly enabling similar Ti isotope analyses of these objects. Indeed, there is increasing evidence for TiO in hot Jupiters from transmission and dayside emission spectra \citep[e.g.,][]{nugroho2017,serindag2021,cont2021,chen2021}. The youngest super Jupiters, still hot from their formation, are also expected to show TiO features, just like brown dwarfs and M-dwarf stars. In fact, TiO band heads are clearly visible in the medium-resolution MUSE optical spectrum of the young, wide-orbiting super-Jupiter GQ Lupi b (Stolker et al., in prep).

In this paper, we use an archival high-dispersion \carmenes{} spectrum of the M-dwarf \gj{} as a proxy for a gas-giant exoplanet observed at extremely high signal-to-noise, and assess the feasibility of performing titanium isotope measurements in exoplanet atmospheres. We investigate how the derived isotope ratios are impacted by different wavelength coverage and loss of continuum information -- a common effect of the processing techniques for high-resolution spectroscopy of exoplanet atmospheres. We also estimate what spectral signal-to-noise ratios are needed to perform such analyses for exoplanets, along with the corresponding exposure times for current 8-m-class telescopes and the future Extremely Large Telescope. In Sect. \ref{sec:data} we describe the high-resolution stellar spectrum we use in this work. We present the high-resolution TiO model spectra in Sect. \ref{sec:model} and outline our fitting methodology in Sect. \ref{sec:mcmc}. We present and discuss the results of our analyses in Sects. \ref{sec:results} and \ref{sec:discussion}, before summarizing our work in Sect. \ref{sec:conclusions}.

\section{High-resolution TiO spectral data} \label{sec:data}

\subsection{\carmenes{} spectrum of \gj{}} \label{sec:reduction}

We use a single high-resolution archival\footnote{Based on data from the \carmenes{} data archive at CAB (INTA-CSIC).} spectrum of the M5.5V \citep{walker1983} star \gj{} taken on 18 November 2016 using the \carmenes{} spectrograph on the 3.5-m Calar Alto Telescope. \carmenes{} \citep{quirrenbach2014} consists of two echelle spectrographs in the visible and near-infrared wavelength regime. The archival observation is a 1238-s exposure in the visual channel, covering 5200--10\,500~\AA{} at a resolving power ${ \mathcal{R} = 94\,600 }$ ($\approx{}$3 \kms{}). The automated \caracal{} pipeline \citep{caballero2016,zechmeister2018} performs the initial data processing including bias correction, order extraction, and wavelength calibration. The archival pipeline product contains the topocentric wavelengths, fluxes, and errors for the 61 orders of the visual channel.

\begin{figure*}
    \centering
    \includegraphics[width=17cm]{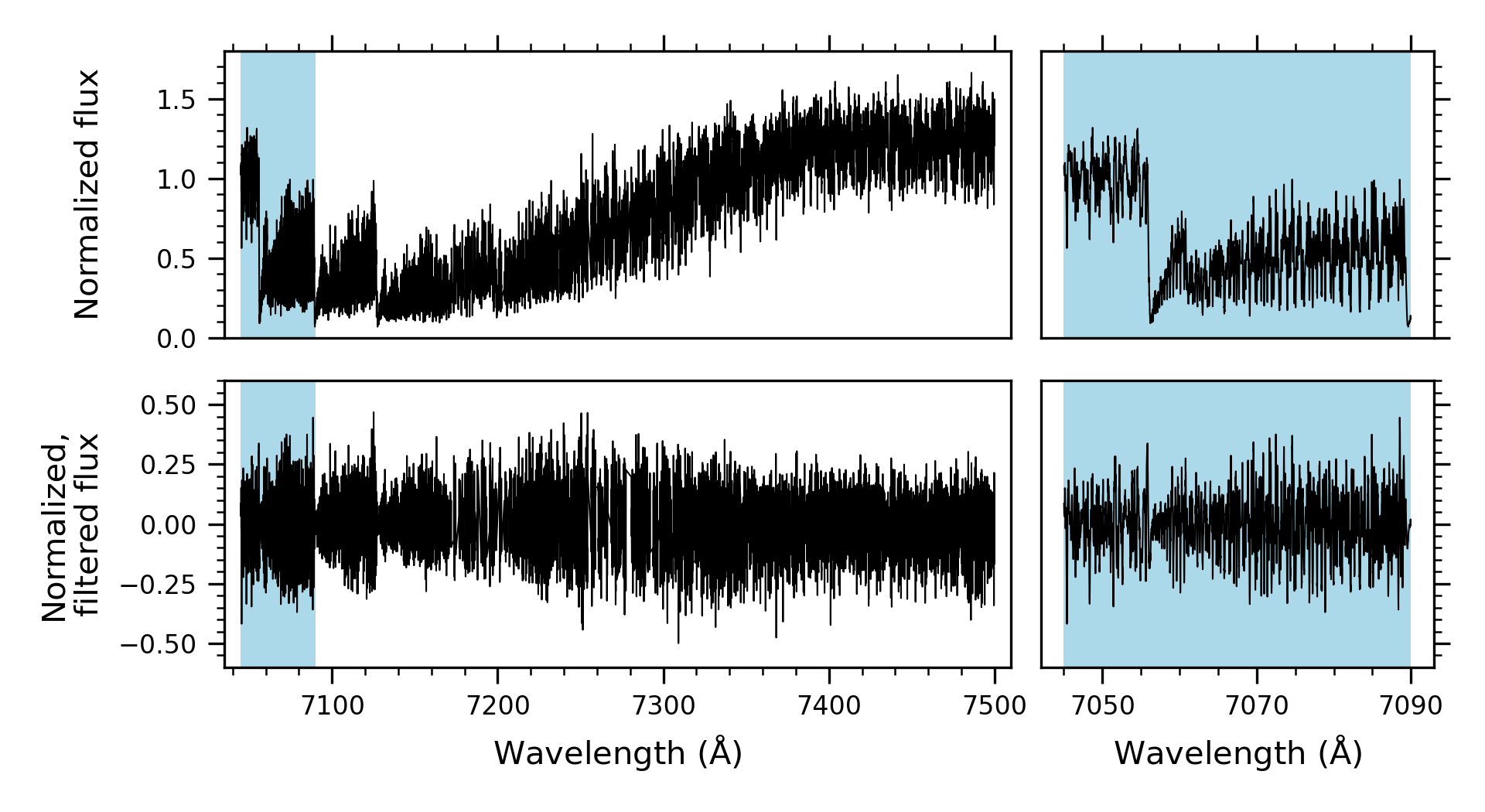}
    \caption{\textit{Top left panel}: Unfiltered \carmenes{} spectrum of \gj{} over the wide wavelength range 7045--7500 \AA{}. The narrow wavelength range 7045-7090 \AA{} is indicated by the blue shading. \textit{Top right panel}: Same, but plotting only the narrow wavelength range for clarity. \textit{Bottom left panel}: Broadband-filtered \carmenes{} spectrum of \gj{} over the wide wavelength range. \textit{Bottom right panel}: Same for the narrow wavelength range. }
    \label{fig:data}
\end{figure*}

We isolate wavelengths 7000--7600 \AA{} from nine orders, and remove all pixels (0.02\%) with non-finite flux values. To mitigate the influence of telluric contamination, we also remove all wavelength bins (9\%) with a telluric transmission value $\le$0.98, based on the ESO SkyCalc model \citep{noll2012,jones2013} with a precipitable water vapor of 2.5 mm. We subsequently shift the spectrum to the stellar rest frame, and divide the flux and error values by the mean flux level in the spectral range 7045--7050 \AA{} preceding the red-degraded $\gamma{}$-system (0,0) band head of TiO at 7054 \AA{}. This wavelength range is similar to those used as continuum levels in previous studies fitting high-resolution stellar TiO features \citep{clegg1979,valenti1998,chavez2009}. The top left panel of Fig. \ref{fig:data} shows the normalized spectrum (hereafter, unfiltered spectrum) over the wavelength range 7045--7500 \AA{}.

To mimic the standard treatment of high-resolution spectra in exoplanet atmosphere analyses, we perform a high-pass filter which results in the loss of continuum information. For each wavelength bin, we subtract the mean flux value in a boxcar with full-width 0.60 \AA{} ($\approx{}$25 \kms{}). Since the pixel spacing is non-uniform in both wavelength and velocity space, we only filter a given pixel if its boxcar contains more than half of the expected number of pixels based on the average wavelength sampling. This results in 2\% of the pixels being excluded. This broadband-filtered spectrum on 7045--7500 \AA{} is shown in the bottom left panel of Fig. \ref{fig:data}.

\subsection{Choice of wavelength range for TiO fitting} \label{sec:spectral_range}

Most of the previous studies that used TiO to determine stellar Ti isotope ratios used the wavelength range $\sim$7050--7100 \AA{}, which contains strong features of the TiO $\gamma{}$-system (0,0) bands. \citet{lambert1972}, \citet{wyckoff1972}, \citet{lambert1977}, and \citet{chavez2009} all chose a range with a relatively low line density between the $\gamma{}$(0,0) band heads at 7054 \AA{} and 7089 \AA{}. \citet{clegg1979} used a similar spectral range, but only for stars with spectral type earlier than M4. For M4 stars and later, they instead analyzed the regions around the $\gamma{}$-system (0,1) band at 7589 \AA{} and $\delta{}$-system (0,0) band around 8900 \AA{}, which are less saturated.

More recently, \citet{pavlenko2020} compared the suitability of various spectral ranges in the optical and near-infrared for TiO isotopologue analysis, and argue that the spectral region 7580--7594 \AA{}, including the $\gamma{}$(0,1) band head at 7589 \AA{}, is preferable to the commonly-used spectral window encompassing the $\gamma{}$(0,0) band. Specifically, they note that while the $\gamma{}$(0,0) band head at 7054 \AA{} is a blend of features from all five TiO isotopologues, the $\gamma{}$(0,1) band heads of \isotope{TiO}{49} and \isotope{TiO}{50} are blue-shifted out of the stronger, red-degraded \isotope{TiO}{48} band head at 7589 \AA{}. Further, they note that the stellar flux for late-type stars is higher at these redder wavelengths than at 7054 \AA{}. This is also expected to be the case for exoplanets. However, a major drawback of using the region surrounding the $\gamma{}$(0,1) band at 7589 \AA{} is the presence of the strong telluric O$_2$ A band at 7590 \AA{}. As \citet{pavlenko2020} point out, the separation between the telluric and TiO band heads -- and thus the level of telluric contamination of the TiO band -- depends on the relative velocity between the target and telescope.

Since the primary objective of this work is to determine the feasibility of deriving accurate Ti isotope ratios from TiO features in the spectra of gas-giant exoplanets, we opt to entirely avoid analyzing spectral regions near the strong O$_2$ A band. Instead, similar to the majority of previous studies, we perform spectral fitting using the range 7045--7090 \AA{} (hereafter, narrow range), shown for the unfiltered and broadband-filtered cases in the top right and bottom right panels of Fig. \ref{fig:data}. To investigate whether including more, albeit weaker, TiO features affects the spectral fitting results for \gj{} or enables more accurate results for noise-degraded cases, we also analyze the spectral range 7045--7500 \AA{} (hereafter, wide range). The top left and bottom left panels of Fig. \ref{fig:data} show the unfiltered and broadband-filtered spectrum, respectively, over this wide wavelength range. For comparison, the narrow range is highlighted in blue.

\section{High-resolution TiO spectral models} \label{sec:model}

We use the radiative transfer code \ptr{} \citep{molliere2019b} to model the TiO spectrum of \gj{}. While originally developed to study exoplanets, the high-resolution emission spectrum functionality of \ptr{} is also appropriate for modeling stellar spectra. We model the atmosphere of \gj{} in 70 layers from 10$^{5}$ to 10$^{-6}$ bar assuming a constant, solar mean molecular weight of 2.33. We linearly interpolate the temperature-pressure (\tp{}) profiles of the MARCS plane-parallel standard-composition stellar model atmospheres grid \citep{gustafsson2008} to the parameters of \gj{}. Specifically, we adopt a microturbulent velocity ${ \turb{} = 2\ \kms{} }$ and fix the remaining stellar parameters based on their literature values from \citet{rajpurohit2018}: ${ \teff{} = 3100\ \K{} }$, ${ \logg{} = +5.5 }$, ${ \metal{} = +0.20 }$. Since the MARCS models do not span a pressure range comparable to our atmospheric model, we subsequently extrapolate the \tp{} profile based on linear fits in $\log{T}$-$\log{P}$ space. To reflect the maximum temperature for which we calculate TiO opacities, we impose a temperature upper limit of 4000 \K{}.

We only include spectroscopic contributions from the five main isotopologues of TiO\footnote{We only differentiate TiO isotopologues based on the stable Ti isotopes. The oxygen isotope fractionation is comparatively negligible.}, as these dominate the M-dwarf optical spectrum. We utilize the recently-released \toto{} line list to calculate TiO opacities. Compared to other commonly-used TiO line lists, \toto{} better reproduces the TiO features in high-resolution M-dwarf spectra for both the main (\isotope{TiO}{48}) and minor isotopologues due to the inclusion of more accurate experimental energy levels in the line list calculations \citep{mcKemmish2019,pavlenko2020}. We use the method outlined in Appendix A of \citet{molliere2015} to calculate TiO opacities up to 4000 \K{} on a high-resolution wavelength grid (${ \lambda{} / d\lambda{} = 10^{6} }$). For all TiO isotopologues, we assume a constant volume mixing ratio (VMR) at each pressure layer in our model atmosphere. 

To facilitate a better fit, we perform similar processing steps on the \ptr{} models as we use on the \carmenes{} data. We broaden the models to the \carmenes{} resolving power and subsequently interpolate them onto the wavelength sampling of the data. We then normalize the model spectra to their mean flux value in the range 7045--7050 \AA{}. If we are fitting the broadband-filtered \carmenes{} spectrum, we also perform a boxcar filtering identical to that described in Sect. \ref{sec:reduction}. As an example, Fig. \ref{fig:iso} plots an unfiltered model (dashed black line) over a limited wavelength range, calculated using TiO isotopologue abundances retrieved for \gj{} (see Sects. 4 and 5). Also shown are the flux contributions of the individual TiO isotopologues (solid lines), demonstrating their influence on the composite model spectrum.

\section{Fitting TiO isotopologue abundances} \label{sec:mcmc}

We use the \emcee{} implementation \citep{dfm2013} of the \citet{goodman2010} affine-invariant Markov chain Monte Carlo ensemble sampler to directly fit the processed \carmenes{} spectrum (Sect. \ref{sec:reduction}) using the similarly-processed \ptr{} models (Sect. \ref{sec:model}). Following the methods presented by \citet{brogi2019} and \citet{gibson2020}, we adopt a log-likelihood function of the form
\begin{equation}
    \lnlike{} = -N \ln{\beta{}} - \frac{1}{2} \sum_{i=1}^{N}{ \left( \frac{ d_i - m_i }{ \beta{} \, \sigma_i  } \right)^2 },
\end{equation}
where $N$ is the number of pixels, $d_i$ and $m_i$ are the data and model flux values for the $i$th pixel, $\sigma{}_i$ is the \carmenes{} error of the $i$th pixel, and $\beta{}$ is a wavelength-invariant scaling factor for the uncertainties. The purpose of the latter term is to allow for the possibility that the \carmenes{} errors are underestimated and to attempt to account for systematic uncertainties in our models.

Our model consists of six fitted parameters: $\beta{}$ and the \lvmr{} for each TiO isotopologue\footnote{For brevity, we denote the \lvmr{} for a given isotopologue \isotope{TiO}{$i$} as \lvmr{i}.}. We adopt uniform priors: ${ [1,\, 100] }$ for $\beta{}$, ${ [-10,\, -6] }$ for \lvmr{48}, and ${ [-13,\, -6] }$ for each minor isotopologue's \lvmr{}. The choice of bounds for the TiO abundances allows for ratios relative to the main isotopologue \isotope{TiO}{48} of ${ [10^{-3},\, 1] }$, which is essentially zero to unity. While we do not fit for \teff{}, \logg{}, \metal{}, \turb{}, and the mean molecular weight (see Sect. \ref{sec:model}), we do perform additional retrievals to estimate the impact uncertainties in these parameters have on the TiO results (see Sect. \ref{sec:discussion_method}).

For each fit, we perform two MCMC runs in sequence. For the first, we initialize fifty walkers uniformly within the bounds of the prior for each fitted parameter, and run the sampler for 500 steps, corresponding to 25\,000 model evaluations. We then initialize fifty walkers in a Gaussian ball around the best-fitting set of parameters from the first run, and evolve the sampler again for 1000 steps (50\,000 model evaluations). We visually inspect and remove the section of the second run preceding convergence. The resulting clipped, converged chain is the set of posterior samples used in our analysis.

\section{Results} \label{sec:results}

\subsection{Ti isotope ratios for the M-dwarf \gj{}} \label{sec:results_noiseless}

\begin{figure*}
    \centering
        \includegraphics[width=17cm]{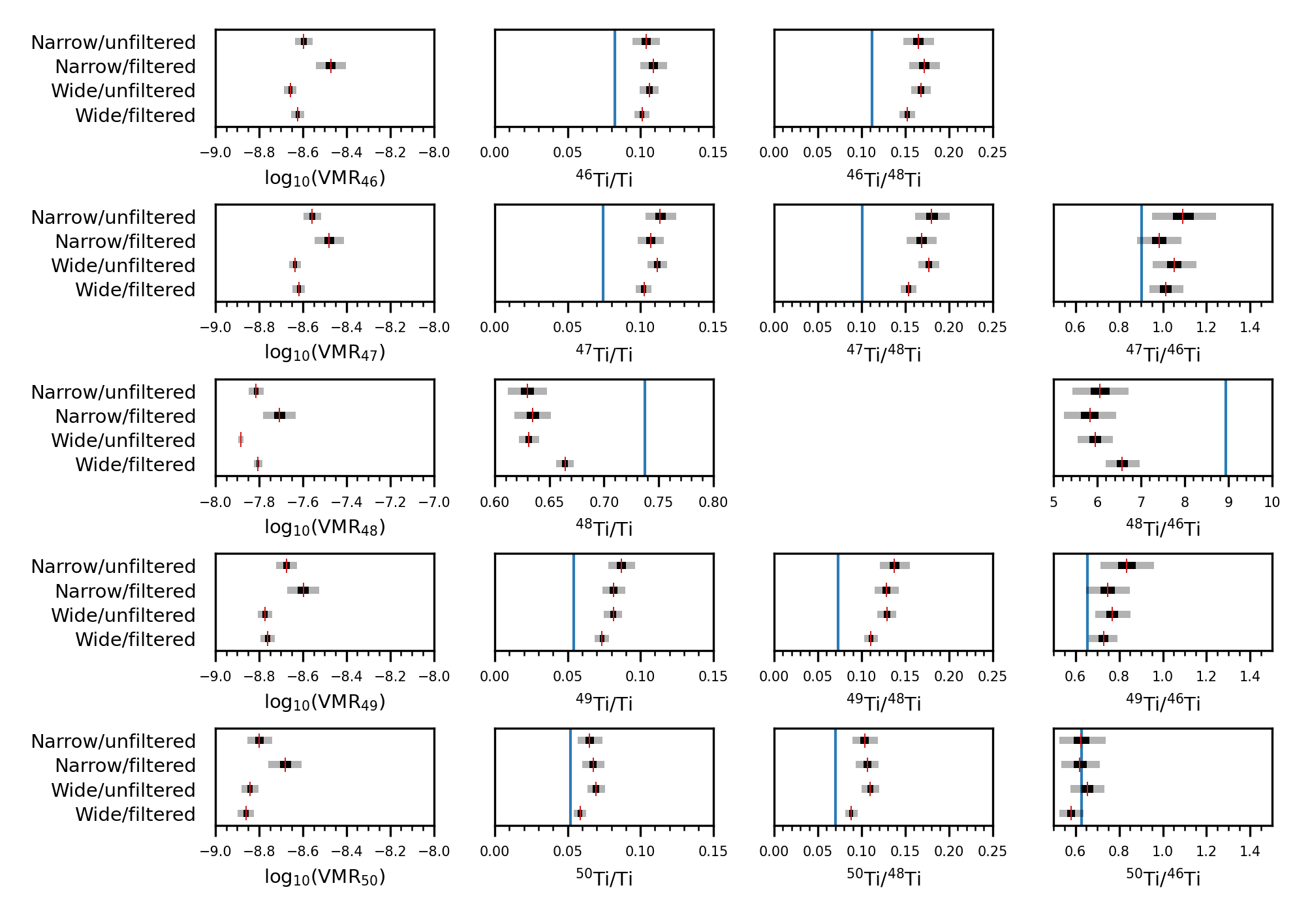}
        \caption{ Results of the MCMC retrievals of TiO features in the \carmenes{} spectrum of \gj{} using the narrow and wide spectral ranges and the unfiltered and broadband-filtered versions of the data. Each panel plots the results for a different parameter, which are grouped in columns. The first column contains panels for the fitted \lvmr{} of the five TiO isotopologues. The second, third, and fourth columns contain panels for the derived isotope ratios relative to Ti, \isotope{Ti}{48}, and \isotope{Ti}{46}, respectively. In each panel, the mean parameter value from each MCMC fitting is indicated by a red marker while the 1\sig{} and 3\sig{} errors are indicated by black and gray bars, respectively. For the panels comparing abundance ratios, we indicate the corresponding terrestrial value with a vertical blue line. }
        \label{fig:results_method_tio}
\end{figure*}

Using the framework presented in Sect. \ref{sec:mcmc}, we perform multiple retrievals of the TiO isotopologue abundances in the M-dwarf \gj{} to assess the impact of different methodologies -- choice of spectral fitting range and broadband filtering. As mentioned in Sect. \ref{sec:spectral_range}, we fit both a narrow (7045--7090 \AA{}) and wide (7045--7500 \AA{}) spectral range to determine the influence of including more TiO lines on the retrieved TiO abundances and Ti isotope ratios. We also perform retrievals on both the unfiltered and broadband-filtered versions of the \carmenes{} data, to assess whether the loss of continuum information inherent to high-resolution studies of exoplanet spectra affects the results. For each methodology -- narrow and unfiltered spectrum, narrow and broadband-filtered spectrum, wide and unfiltered spectrum, wide and broadband-filtered spectrum -- we perform three independent retrievals to demonstrate consistency, leading to a total of twelve MCMC retrievals. In all twelve cases, the second MCMC run provides a uni-modal solution with constrained posteriors for all six fitted parameters.

The first column of Fig. \ref{fig:results_method_tio} plots the fitted TiO abundances for the various MCMC retrievals. Each panel corresponds to the \lvmr{} for a different TiO isotopologue, with the mean posterior value for a given fit shown by the red marker and the 1\sig{} and 3\sig{} errors indicated by the black and gray bars, respectively. For reference, the mean values and 1\sig{} errors for all fitted parameters -- including $\beta{}$ -- are given Table \ref{table:method_parameters}. We also derive sets of posteriors for the isotope ratios relative to Ti, \isotope{Ti}{48}, and \isotope{Ti}{46}. The corresponding mean and 1\sig{} errors are also provided in Table \ref{table:method_parameters}, and plotted in the second, third, and fourth columns of Fig. \ref{fig:results_method_tio}. We only provide results of the first retrieval for each methodology in Fig. \ref{fig:results_method_tio} and Table \ref{table:method_parameters} because the second and third retrievals give very similar values.

\subsection{Ti isotope ratios from noise-degraded spectra} \label{sec:addedNoiseResults}

\begin{figure*}
    \centering
        \includegraphics[width=17cm]{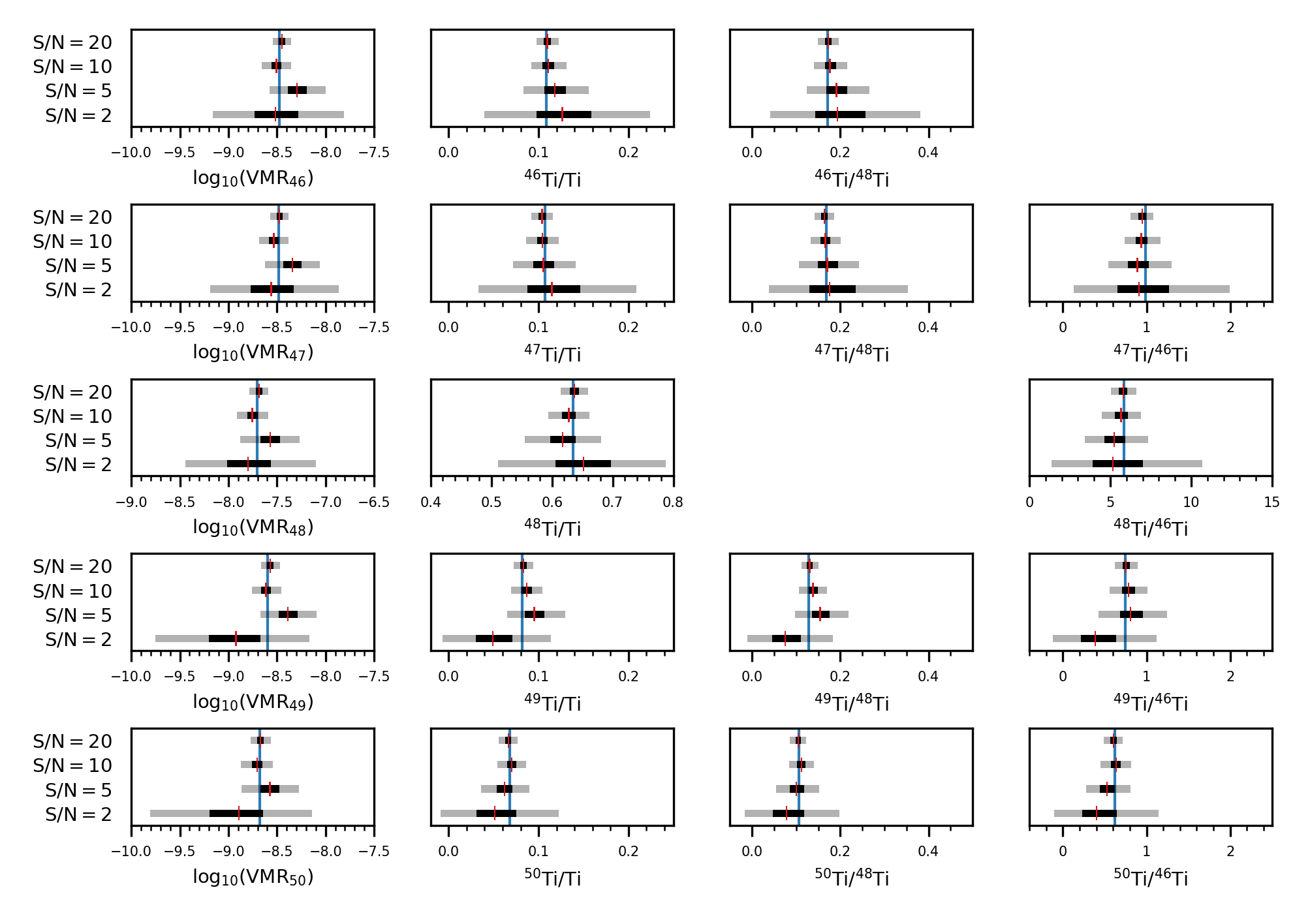}
        \caption{ Same as Fig. \ref{fig:results_method_tio}, but for the MCMC retrievals of TiO features in the narrow, broadband-filtered \carmenes{} spectrum of \gj{} with various levels of noise added. The results from the retrieval of the narrow/broadband-filtered spectrum without added noise are shown by the vertical blue lines.}
        \label{fig:results_noise_tio_nf}
\end{figure*}

\begin{figure*}
    \centering
        \includegraphics[width=17cm]{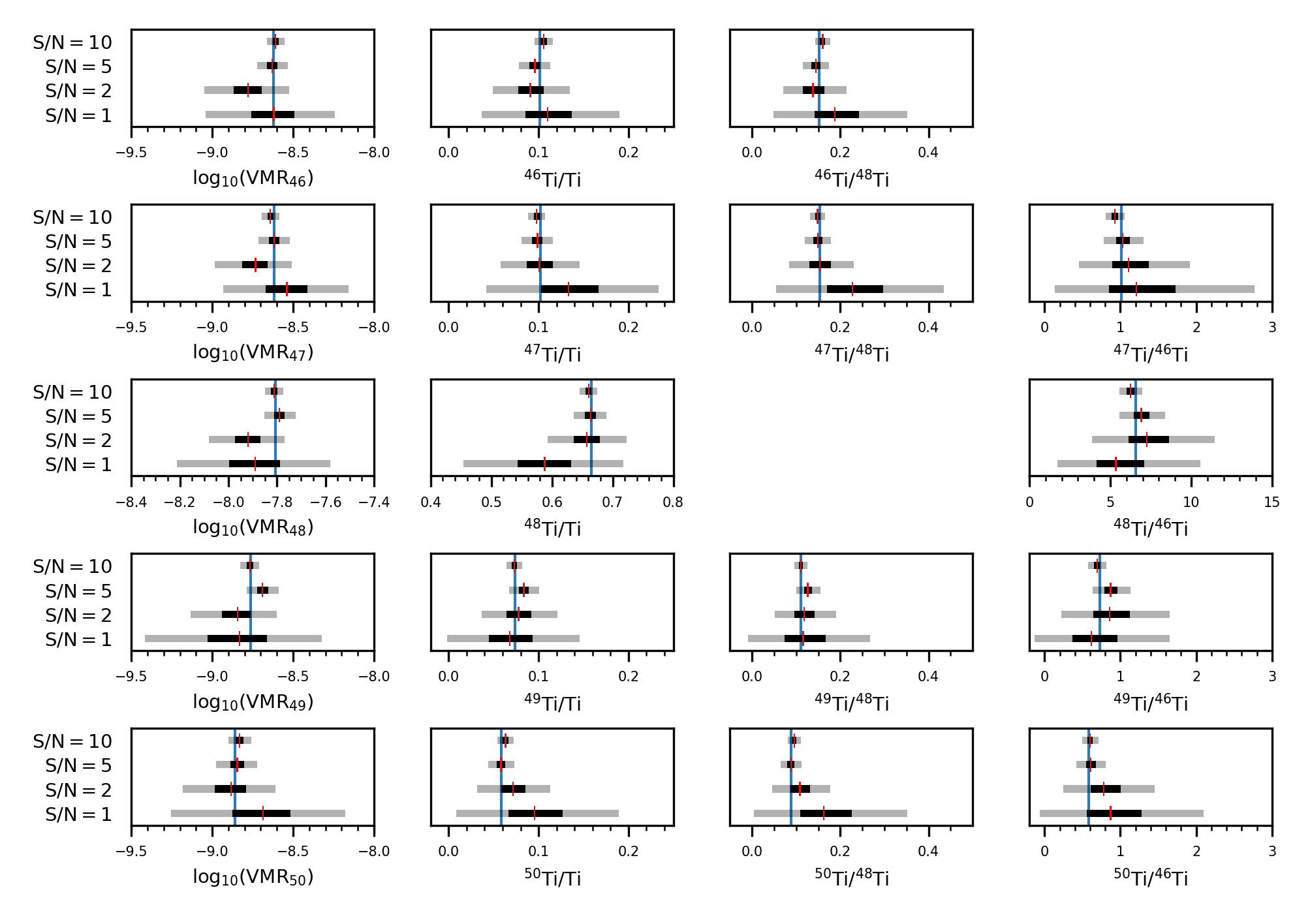}
        \caption{ Same as Fig. \ref{fig:results_noise_tio_nf}, but for the MCMC retrievals performed on the wide, broadband-filtered \carmenes{} spectrum of \gj{}. }
        \label{fig:results_noise_tio_wf}
\end{figure*}

To assess the effectiveness of this technique for deriving Ti isotope ratios in the gas-giant exoplanet case, we degrade the \carmenes{} spectrum of \gj{} with varying levels of noise and re-run the MCMC fitting. To achieve a given planet \snr{} on the wavelength range used for normalization (7045--7050 \AA{}), we add white noise to the pipeline-reduced \carmenes{} spectrum by sampling a zero-mean normal distribution with standard deviation equal to the average flux value on 7045--7050 \AA{} divided by the desired \snr{}. We subsequently perform the same telluric-contaminated pixel removal, reference frame shift, normalization, and high-pass filtering described in Sect. \ref{sec:reduction}. We then use the same MCMC retrieval procedure described in Sect. \ref{sec:mcmc} to fit the TiO isotopologue features for both the narrow and wide wavelength ranges in each noise-degraded, broadband-filtered data set. As before, we run each fitting three times to ensure consistency of results.

Similar to Fig. \ref{fig:results_method_tio}, we present the fitted TiO isotopologue abundance values for each noise case in the first columns of Figs. \ref{fig:results_noise_tio_nf} and \ref{fig:results_noise_tio_wf} for the narrow and broadband-filtered spectrum, and the wide and broadband-filtered spectrum, respectively. The second, third, and fourth columns of these Figs. show the derived isotope ratios relative to Ti, \isotope{Ti}{48}, and \isotope{Ti}{46}, respectively. For the narrow/broadband-filtered case, we present results for spectra degraded to \snr{} of 20, 10, 5, and 2, while for the wide/broadband-filtered case we show the results for spectra degraded to \snr{} of 10, 5, 2, and 1. Again, because the results of the three retrievals for each noise case are very similar, we only provide the values from the first retrieval in Figs. \ref{fig:results_noise_tio_nf} and \ref{fig:results_noise_tio_wf}.

\section{Discussion} \label{sec:discussion}

\begin{figure*}
    \centering
        \includegraphics[width=17cm]{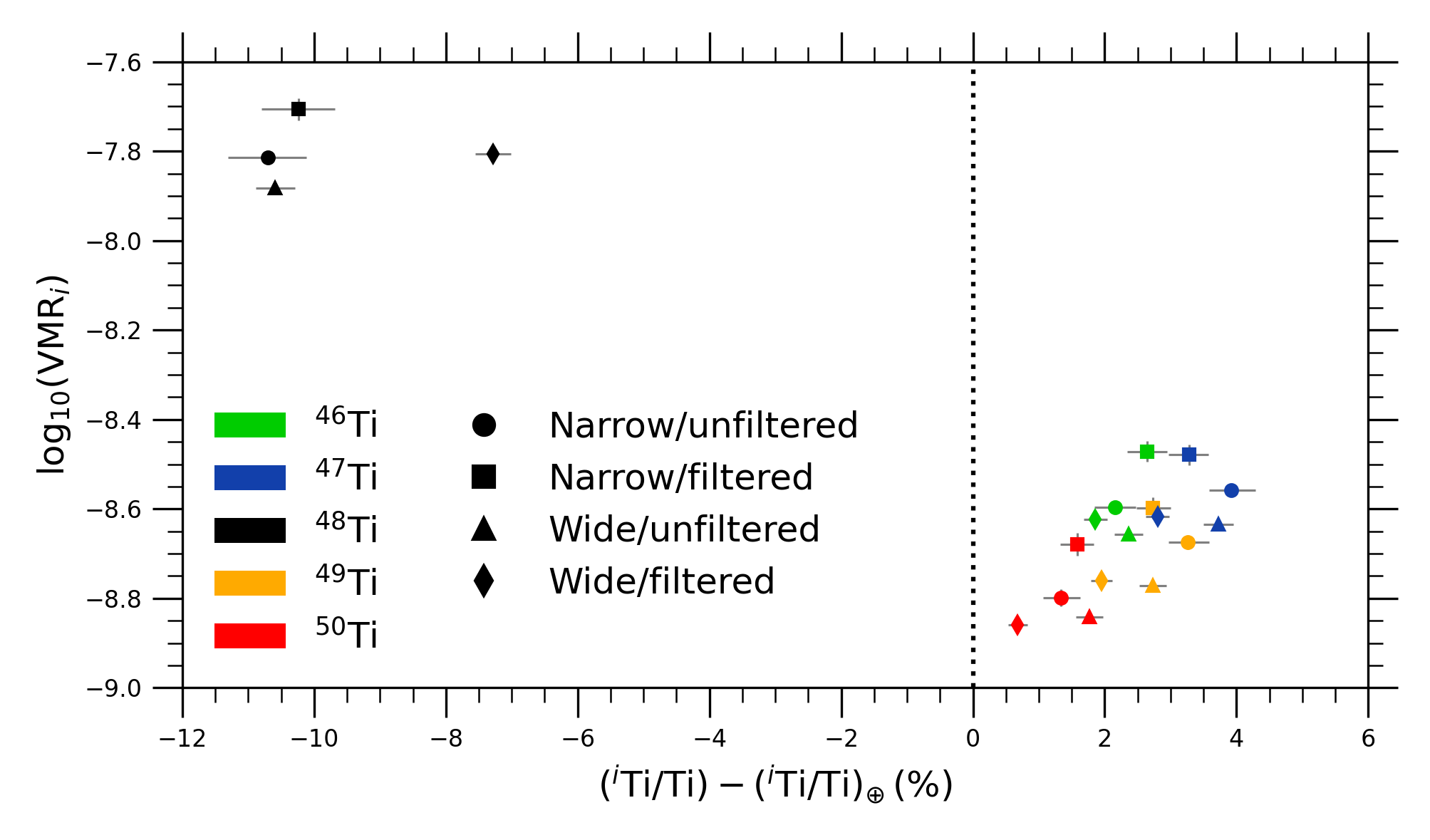}
        \caption{ Comparison of absolute and relative abundances derived from each retrieval methodology, demonstrating the small impact wavelength range and broadband filtering have on the results. For each stable titanium isotope \isotope{Ti}{$i$}, the retrieved \lvmr{} of the corresponding isotopologue \isotope{TiO}{$i$} is shown on the $y$-axis, while the $x$-axis shows the absolute deviation in the relative abundance \isotope{Ti}{$i$}/\isotope{Ti}{} compared to the terrestrial value (vertical dotted line). The markers colored green, blue, black, orange, and red are used for \isotope{Ti}{46}, \isotope{Ti}{47}, \isotope{Ti}{48}, \isotope{Ti}{49} and \isotope{Ti}{50}, respectively.  The results for the narrow and unfiltered, narrow and filtered, wide and unfiltered, and wide and filtered spectra are denoted by the circular, square, triangular, and diamond markers, respectively. 1\sig{} errors, derived from the MCMC posteriors, are shown in gray.}
        \label{fig:2d_deviation}
\end{figure*}

\subsection{Effects of wavelength range and broadband filtering} \label{sec:discussion_method}

Small systematic effects ($\lesssim$0.1 dex) seem to be present between isotopologue abundances retrieved using the narrow and wide wavelength ranges, with the latter resulting in lower \lvmr{} values for a given filtering treatment (see Fig. \ref{fig:results_method_tio}, first column). These effects likely arise from the additional sources of TiO opacity contained on the wider spectral range. For instance, in addition to the red-degraded $\gamma{}$(0,0) band starting at 7054 \AA{}, the wide-spectrum retrievals also contain the two other strong red-degraded bands of the TiO $\gamma{}$(0,0) triplet with band heads at 7089 \AA{} and 7126 \AA{}, as well as additional weak lines further red-wards. Simultaneously probing more features of varying strengths arising from different locations in the atmosphere could well lead to different retrieved abundances. Similarly-small variations in abundance are also seen between retrievals using the unfiltered and broadband-filtered spectra. For a given wavelength range, fitting the broadband-filtered data generally leads to higher \lvmr{} values. These differences can be attributed to the loss of continuum information, but are only $\sim$0.01 dex for the wide spectral range, except for the main isotopologue ($\sim$0.1 dex). Thus, the loss of continuum information hardly affects the retrieval of isotopologue abundances.

The remaining columns of Fig. \ref{fig:results_method_tio}, comparing the derived Ti isotope ratios relative to Ti, \isotope{Ti}{48}, and \isotope{Ti}{46}, show there is generally good agreement between different retrievals, particularly for those ratios relative to \isotope{Ti}{46}. In Table \ref{table:method_agreement}, we provide the formal deviation between parameters retrieved using the different methodologies, calculated based on the errors derived from the MCMC posteriors. We note that despite our inclusion of the $\beta{}$ parameter in the log-likelihood function (see Sect. \ref{sec:mcmc}), these formal deviation values are likely conservative error estimates and do not fully account for various sources of systematic uncertainty, for instance, in the choice of \tp{} profile, line list inaccuracies, or spectral modeling technique. Across all four retrieval methods, the isotope ratios relative to \isotope{Ti}{46} deviate by $<$3\sig{} for the minor isotopes and $<$4\sig{} for \isotope{Ti}{48}. This greater deviation for \isotope{Ti}{48}/\isotope{Ti}{46} may be due to saturation of the strongest \isotope{TiO}{48} lines, as noted in previous studies \citep{clegg1979,chavez2009,pavlenko2020}. The absolute spread in isotope ratios relative to Ti and \isotope{Ti}{48} across the various retrieval setups is only $\sim$1\%. This small variation is immediately evident in Fig. \ref{fig:2d_deviation}, where we plot the absolute and relative abundances derived using the different methodologies (denoted by different markers), and find rather close clustering of values for each Ti isotope (denoted by different colors). This demonstrates that, like the retrieved isotopologue abundances, the isotope ratios relative to Ti and \isotope{Ti}{48} do not deviate much in absolute terms due to loss of continuum information or choice of wavelength range. This is particularly important considering the that the signal-to-noise ratio of exoplanet spectra will always be significantly lower than the stellar spectrum used here, resulting in much larger random errors (see Sect. \ref{sec:discussion_noise}).

We also verify that our assumptions for the mean molecular weight $\mu$ and stellar parameters of the \tp{} profile do not substantially impact the Ti isotope analysis. For both the narrow and broadband-filtered spectrum and the wide and broadband-filtered spectrum, we perform MCMC fittings with the following mono-substituted parameters: ${ \mu = \{1.9,\ 2.1\} }$, ${ \teff{} = \{3000,\ 3200\} \ \mathrm{K} }$, ${ \logg{} = +5.0 }$, ${ \metal{} = \{+0.00,\ +0.50\} }$, and ${ \turb{} = 1 \, \kms{} }$. Varying $\mu$ has negligible impact on the retrieved isotope ratios. Changing the stellar \tp{} parameters results in $<$3\sig{} deviations for all isotope ratios relative to \isotope{Ti}{46} and most isotope ratios relative to Ti and \isotope{Ti}{48}.

\subsection{\gj{} Ti isotope ratios in context} \label{sec:discussion_context}

The relative abundances of Ti isotopes in \gj{} derived using the four different retrieval setups are all significantly different from the terrestrial values. While \isotope{Ti}{48} accounts for about 74\% of all titanium atoms on Earth, we find this value to be only 63--66\% in \gj{}. The minor isotopes are all measured to be about 1-4\% more abundant. We compare these values to those measured by \citet{chavez2009} for a sample of eleven K- and M-dwarfs. They find a spread of about 11\% in the relative abundance of \isotope{Ti}{48}, which is similar to the difference we find between \gj{} and Earth. With a literature metallicity of ${ \metal{} = +0.20 }$, the lower \isotope{Ti}{48}/\isotope{Ti}{} value we measure for \gj{} is consistent with expectations from various Galactic chemical models \citep[e.g.,][]{hughes2008,chavez2009}, which predict a negative correlation between the relative abundance of \isotope{Ti}{48} and metallicity. However, we note that in their stellar sample spanning metallicities $-1 < \mathrm{[Fe/H]} < 0$, \citet{chavez2009} find essentially no correlation between metallicity and isotope ratios relative to \isotope{Ti}{48}, which are all similar to terrestrial.

As mentioned in Sect. \ref{sec:discussion_method}, though, it is likely the strong \isotope{TiO}{48} lines we fit are saturated, so we are cautious to draw firm conclusions from the isotope ratios relative to Ti and \isotope{Ti}{48}. On the other hand, the abundances ratios of \isotope{Ti}{47}, \isotope{Ti}{49}, and \isotope{Ti}{50} relative to \isotope{Ti}{46} are not expected to be affected by saturation. The fourth column of Fig. \ref{fig:results_method_tio} indicates \isotope{Ti}{50}/\isotope{Ti}{46} in \gj{} to be consistent with the terrestrial value, and the values for \isotope{Ti}{47}/\isotope{Ti}{46} and \isotope{Ti}{49}/\isotope{Ti}{46} to be consistent or marginally higher than Earth. These results are generally consistent with those from \citet{chavez2009}, who find approximately terrestrial values across their M-dwarf sample.

\subsection{Determining Ti isotope ratios for gas-giant exoplanets} \label{sec:discussion_noise}

In Sect. \ref{sec:discussion_method}, we show that the loss of continuum information common to exoplanet studies at high resolution should not be prohibitive in deriving Ti isotope ratios for gas-giant exoplanets. The other major difference between TiO analyses of M-dwarfs and future studies of exoplanets is the much fainter signals in the latter case. Figures \ref{fig:results_noise_tio_nf} and \ref{fig:results_noise_tio_wf} demonstrate that even for relatively low \snr{} levels, it is possible to constrain the relative abundances of Ti isotopes from the narrow and broadband-filtered spectrum and the wide and broadband-filtered spectrum, respectively. For each of the \snr{} levels shown, the \lvmr{} value of each isotopologue is consistent to within 3\sig{} of the non-degraded results, as are the Ti abundance ratios.

As expected, the fitting errors from the MCMC posteriors increase as the \snr{} decreases, and are smaller for the wide/broadband-filtered spectrum compared to those for the narrow/broadband-filtered spectrum at a given \snr{}. To achieve $\lesssim$10\% relative error for all Ti isotope ratios, the spectra in the range 7045-7050 \AA{} must have ${ \snr{} \ge 10 }$ for the narrow/broadband-filtered case and ${ \snr{} \ge 5 }$ for the wide/broadband-filtered case.

To estimate the integration time required to achieve these \snr{} for various planetary systems using future high-contrast/high-dispersion instruments such as RISTRETTO \citep{chazelas2020} on the Very Large Telescope (VLT) and HIRES \citep{marconi2020} on the Extremely Large Telescope (ELT), we use a method similar to that described in \citet{molliere2019a}. Assuming Poisson noise and negligible sky contribution, the planet emission \snr{} in a pixel centered at wavelength $\lambda_0$ is
\begin{equation} \label{eq:snr_planet}
    \snrP{} = \frac{\contrast{}}{\sqrt{ \contrast{} + 1/f }} \snrS{},
\end{equation}
where \snrS{} is the stellar \snr{} in the pixel, \contrast{} is the planet-to-star luminosity contrast at wavelength $\lambda_0$, and $f$ is the stellar-flux-reduction factor used to estimate the effect of suppressed stellar contribution in spatially-resolved observations. Following \citet{molliere2019a}, we adopt $f$ values of 100 and 1000 for spatially-resolved observations on the VLT and ELT, respectively, and ${f=1}$ for spatially-unresolved observations. We approximate \contrast{} as the ratio of blackbody luminosities at $\lambda_0$.

The stellar \snr{} is related to the integration time \tint{} by
\begin{equation} \label{eq:snr_star}
    \snrS{} =   \sqrt{\nphotS{}} =
                \sqrt{\frac{ \fluxSE{}\ \area{}\ \tint{}\ \dlpix{}\ \tau{} }{ \ephot{} }},
\end{equation}
where \nphotS{} is the number of stellar photons collected by the pixel centered at $\lambda_0$, \fluxSE{} is the stellar flux at wavelength $\lambda_0$ received at Earth, \area{} is the telescope collecting area, \dlpix{} is the pixel width, $\tau{}$ is the telescope/instrument throughput, and \ephot{} is the photon energy at wavelength $\lambda_0$. Like \citet{molliere2019a}, we take \area{} to be 52 m$^2$ and 976 m$^2$ for the VLT and ELT, respectively, and adopt a throughput of 0.15. We approximate \fluxSE{} using the blackbody luminosity at wavelength $\lambda_0$. For a given resolving power and number of pixels per resolution element \npix{}, we can write ${ \dlpix{} = \lambda_0/ (\mathcal{R}\ \npix{}) }$. Following \citet{molliere2019a}, we adopt ${\mathcal{R}=100\,000}$ and ${\npix{}=3}$. Replacing for \dlpix{}, equating Eqs. (\ref{eq:snr_planet}) and (\ref{eq:snr_star}), and rearranging, the required integration time for a given planet \snr{} is
\begin{equation} \label{eq:tint}
    \tint{} =   \left[\snrP{}\right]^{2}
                \left(\frac{ \ephot{}\ \mathcal{R}\ \npix{} }{ \fluxSE{}\ \area{}\ \tau{}\  \lambda_0 } \right)
                \left(\frac{\contrast{} + 1/f}{\contrast{}^2}\right) .
\end{equation}

Based on Eq. (\ref{eq:tint}), deriving Ti isotope ratios in unresolved (${f=1}$) observations of hot Jupiters is possible, but challenging. Consider, for instance, the WASP-33 system, consisting of the transiting 1.7-\rj{} ultra-hot-Jupiter WASP-33b (${T_\mathrm{day} = 3100 \, \mathrm{K} }$) orbiting its 1.5-\rs{} A5 host star (${\teff{} = 7430 \, \mathrm{K}}$) in a 1.2-d period \citep{cameron2010,kovacs2013,nugroho2021}. WASP-33b is, thus far, the only hot Jupiter with evidence for TiO emission based on high-resolution spectral analyses \citep{nugroho2017,serindag2021,cont2021}. At a distance of 117 pc \citep{kovacs2013}, to achieve ${\snr{} = 5}$ at 7045 \AA{} on the ELT requires about seven hours of integration time assuming the planet dayside is fully visible, but nearly 29 hours of integration time assuming only half the planet dayside is visible.

Spatially-resolved observations decrease the required integration time by reducing stellar contamination, though the planets suitable to such observations necessarily orbit much farther from their host stars than hot Jupiters. However, young wide-orbit planets are known to have \teff{} similar to ultra-hot Jupiters. For example, the directly-imaged 3.4-\rj{} planet GQ Lupi b has an effective temperature of 2400 K, despite orbiting its 1.7-\rs{} T Tauri K7 host star (${\teff{} = 4300 \, \mathrm{K}}$) at $\sim$100 AU \citep{herbig1977,lavigne2009,donati2012,wu2017}. At 150 pc \citep{crawford2000} it would take less than one minute of integration time on the ELT to achieve a planet \snr{} of 5 at 7045 \AA{}. Excitingly, the same \snr{} would also be attainable on the VLT in about an hour. Thus, there are excellent prospects to determine Ti isotope ratios in young gas-giant exoplanets on wide orbits.

\section{Conclusions} \label{sec:conclusions}

We used \ptr{} models to fit TiO features in a \carmenes{} high-resolution spectrum of the M-dwarf \gj{}, retrieving the relative abundances of Ti isotopes for a narrow and wide wavelength range, and for the unfiltered and broadband-filtered spectrum. The latter mimics typical high-resolution spectra of exoplanets for which continuum information is lost. Differences in the retrieved isotope ratios using the different setups are small. Most affected is the main isotope \isotope{Ti}{48}, due to possible saturation of the strongest \isotope{TiO}{48} lines. By degrading the \snr{} of the \gj{} spectrum and re-running the retrievals, we determine a planetary ${\snr{} \ge 5}$ is necessary to retrieve abundance ratios with relative errors ${\lesssim}$10\% when fitting the wide wavelength range 7045--7500 \AA{}. Future spatially-resolved high-dispersion observations of wide-orbiting young gas giants can easily achieve such \snr{}, requiring only an hour and less than a minute of integration time on VLT/RISTRETTO and ELT/HIRES, respectively.

\begin{acknowledgements}

DBS, IAGS, and PM acknowledge support from the European Research Council under the European Union’s Horizon 2020 research and innovation program under grant agreement No. 694513.

PM acknowledges support from the European Research Council under the European Union’s Horizon 2020 research and innovation program under grant agreement No. 832428.

\end{acknowledgements}


\bibliographystyle{aa}
\bibliography{tio_iso.bib}

\begin{thebibliography}{58}
\expandafter\ifx\csname natexlab\endcsname\relax\def\natexlab#1{#1}\fi

\bibitem[{{Altwegg} {et~al.}(2015){Altwegg}, {Balsiger}, {Bar-Nun},
  {Berthelier}, {Bieler}, {Bochsler}, {Briois}, {Calmonte}, {Combi}, {De
  Keyser}, {Eberhardt}, {Fiethe}, {Fuselier}, {Gasc}, {Gombosi}, {Hansen},
  {H{\"a}ssig}, {J{\"a}ckel}, {Kopp}, {Korth}, {LeRoy}, {Mall}, {Marty},
  {Mousis}, {Neefs}, {Owen}, {R{\`e}me}, {Rubin}, {S{\'e}mon}, {Tzou}, {Waite},
  \& {Wurz}}]{altwegg2015}
{Altwegg}, K., {Balsiger}, H., {Bar-Nun}, A., {et~al.} 2015, Science, 347,
  1261952

\bibitem[{{Asplund} {et~al.}(2009){Asplund}, {Grevesse}, {Sauval}, \&
  {Scott}}]{asplund2009}
{Asplund}, M., {Grevesse}, N., {Sauval}, A.~J., \& {Scott}, P. 2009, \araa, 47,
  481

\bibitem[{{Ayres} {et~al.}(2013){Ayres}, {Lyons}, {Ludwig}, {Caffau}, \&
  {Wedemeyer-B{\"o}hm}}]{ayres2013}
{Ayres}, T.~R., {Lyons}, J.~R., {Ludwig}, H.~G., {Caffau}, E., \&
  {Wedemeyer-B{\"o}hm}, S. 2013, \apj, 765, 46

\bibitem[{{Brogi} \& {Line}(2019)}]{brogi2019}
{Brogi}, M. \& {Line}, M.~R. 2019, \aj, 157, 114

\bibitem[{Caballero {et~al.}(2016)Caballero, Guàrdia, del Fresno, Zechmeister,
  de~Juan, Alonso-Floriano, Amado, Colomé, Cortés-Contreras, García-Piquer,
  Gesa, de~Guindos, Hagen, Helmling, Castaño, Kürster, López-Santiago,
  Montes, Muñoz, Pavlov, Quirrenbach, Reiners, Ribas, Seifert, \&
  Solano}]{caballero2016}
Caballero, J.~A., Guàrdia, J., del Fresno, M.~L., {et~al.} 2016, in
  Observatory Operations: Strategies, Processes, and Systems VI, ed. A.~B.
  Peck, R.~L. Seaman, \& C.~R. Benn, Vol. 9910, International Society for
  Optics and Photonics (SPIE), 110 -- 127

\bibitem[{Chavez \& Lambert(2009)}]{chavez2009}
Chavez, J. \& Lambert, D.~L. 2009, \apj, 699, 1906

\bibitem[{Chazelas {et~al.}(2020)Chazelas, Lovis, Blind, Kühn, Genolet,
  Hughes, Turbet, Hagelberg, Restori, Kasper, \& Urra}]{chazelas2020}
Chazelas, B., Lovis, C., Blind, N., {et~al.} 2020, in Adaptive Optics Systems
  VII, ed. L.~Schreiber, D.~Schmidt, \& E.~Vernet, Vol. 11448, International
  Society for Optics and Photonics (SPIE), 1393 -- 1401

\bibitem[{{Chen} {et~al.}(2021){Chen}, {Pall{\'e}}, {Parviainen}, {Murgas}, \&
  {Yan}}]{chen2021}
{Chen}, G., {Pall{\'e}}, E., {Parviainen}, H., {Murgas}, F., \& {Yan}, F. 2021,
  \apjl, 913, L16

\bibitem[{{Clegg} {et~al.}(1979){Clegg}, {Lambert}, \& {Bell}}]{clegg1979}
{Clegg}, R.~E.~S., {Lambert}, D.~L., \& {Bell}, R.~A. 1979, \apj, 234, 188

\bibitem[{{Collier Cameron} {et~al.}(2010){Collier Cameron}, {Guenther},
  {Smalley}, {McDonald}, {Hebb}, {Andersen}, {Augusteijn}, {Barros}, {Brown},
  {Cochran}, {Endl}, {Fossey}, {Hartmann}, {Maxted}, {Pollacco}, {Skillen},
  {Telting}, {Waldmann}, \& {West}}]{cameron2010}
{Collier Cameron}, A., {Guenther}, E., {Smalley}, B., {et~al.} 2010, \mnras,
  407, 507

\bibitem[{{Cont} {et~al.}(2021){Cont}, {Yan}, {Reiners}, {Casasayas-Barris},
  {Molli{\`e}re}, {Pall{\'e}}, {Henning}, {Nortmann}, {Stangret}, {Czesla},
  {L{\'o}pez-Puertas}, {S{\'a}nchez-L{\'o}pez}, {Rodler}, {Ribas},
  {Quirrenbach}, {Caballero}, {Amado}, {Carone}, {Khaimova}, {Kreidberg},
  {Molaverdikhani}, {Montes}, {Morello}, {Nagel}, {Oshagh}, \&
  {Zechmeister}}]{cont2021}
{Cont}, D., {Yan}, F., {Reiners}, A., {et~al.} 2021, \aap, 651, A33

\bibitem[{{Crawford}(2000)}]{crawford2000}
{Crawford}, I.~A. 2000, \mnras, 317, 996

\bibitem[{{Donati} {et~al.}(2012){Donati}, {Gregory}, {Alencar}, {Hussain},
  {Bouvier}, {Dougados}, {Jardine}, {M{\'e}nard}, \& {Romanova}}]{donati2012}
{Donati}, J.~F., {Gregory}, S.~G., {Alencar}, S.~H.~P., {et~al.} 2012, \mnras,
  425, 2948

\bibitem[{{Foreman-Mackey} {et~al.}(2013){Foreman-Mackey}, {Hogg}, {Lang}, \&
  {Goodman}}]{dfm2013}
{Foreman-Mackey}, D., {Hogg}, D.~W., {Lang}, D., \& {Goodman}, J. 2013, \pasp,
  125, 306

\bibitem[{{Fortney} {et~al.}(2008){Fortney}, {Lodders}, {Marley}, \&
  {Freedman}}]{fortney2008}
{Fortney}, J.~J., {Lodders}, K., {Marley}, M.~S., \& {Freedman}, R.~S. 2008,
  \apj, 678, 1419

\bibitem[{{Gandhi} \& {Madhusudhan}(2019)}]{gandhi2019}
{Gandhi}, S. \& {Madhusudhan}, N. 2019, \mnras, 485, 5817

\bibitem[{{Genda} \& {Ikoma}(2008)}]{genda2008}
{Genda}, H. \& {Ikoma}, M. 2008, \icarus, 194, 42

\bibitem[{{Gibson} {et~al.}(2020){Gibson}, {Merritt}, {Nugroho}, {Cubillos},
  {de Mooij}, {Mikal-Evans}, {Fossati}, {Lothringer}, {Nikolov}, {Sing},
  {Spake}, {Watson}, \& {Wilson}}]{gibson2020}
{Gibson}, N.~P., {Merritt}, S., {Nugroho}, S.~K., {et~al.} 2020, \mnras, 493,
  2215

\bibitem[{{Goodman} \& {Weare}(2010)}]{goodman2010}
{Goodman}, J. \& {Weare}, J. 2010, Communications in Applied Mathematics and
  Computational Science, 5, 65

\bibitem[{{Gustafsson} {et~al.}(2008){Gustafsson}, {Edvardsson}, {Eriksson},
  {J{\o}rgensen}, {Nordlund}, \& {Plez}}]{gustafsson2008}
{Gustafsson}, B., {Edvardsson}, B., {Eriksson}, K., {et~al.} 2008, \aap, 486,
  951

\bibitem[{{Hartogh} {et~al.}(2011){Hartogh}, {Lis}, {Bockel{\'e}e-Morvan}, {de
  Val-Borro}, {Biver}, {K{\"u}ppers}, {Emprechtinger}, {Bergin}, {Crovisier},
  {Rengel}, {Moreno}, {Szutowicz}, \& {Blake}}]{hartogh2011}
{Hartogh}, P., {Lis}, D.~C., {Bockel{\'e}e-Morvan}, D., {et~al.} 2011, \nat,
  478, 218

\bibitem[{{Herbig}(1977)}]{herbig1977}
{Herbig}, G.~H. 1977, \apj, 214, 747

\bibitem[{{Hubeny} {et~al.}(2003){Hubeny}, {Burrows}, \&
  {Sudarsky}}]{hubeny2003}
{Hubeny}, I., {Burrows}, A., \& {Sudarsky}, D. 2003, \apj, 594, 1011

\bibitem[{{Hughes} {et~al.}(2008){Hughes}, {Gibson}, {Carigi},
  {S{\'a}nchez-Bl{\'a}zquez}, {Chavez}, \& {Lambert}}]{hughes2008}
{Hughes}, G.~L., {Gibson}, B.~K., {Carigi}, L., {et~al.} 2008, \mnras, 390,
  1710

\bibitem[{{Jones} {et~al.}(2013){Jones}, {Noll}, {Kausch}, {Szyszka}, \&
  {Kimeswenger}}]{jones2013}
{Jones}, A., {Noll}, S., {Kausch}, W., {Szyszka}, C., \& {Kimeswenger}, S.
  2013, \aap, 560, A91

\bibitem[{{Kov{\'a}cs} {et~al.}(2013){Kov{\'a}cs}, {Kov{\'a}cs}, {Hartman},
  {Bakos}, {Bieryla}, {Latham}, {Noyes}, {Reg{\'a}ly}, \&
  {Esquerdo}}]{kovacs2013}
{Kov{\'a}cs}, G., {Kov{\'a}cs}, T., {Hartman}, J.~D., {et~al.} 2013, \aap, 553,
  A44

\bibitem[{{Lambert} \& {Luck}(1977)}]{lambert1977}
{Lambert}, D.~L. \& {Luck}, R.~E. 1977, \apj, 211, 443

\bibitem[{{Lambert} \& {Mallia}(1972)}]{lambert1972}
{Lambert}, D.~L. \& {Mallia}, E.~A. 1972, \mnras, 156, 337

\bibitem[{{Lavigne} {et~al.}(2009){Lavigne}, {Doyon}, {Lafreni{\`e}re},
  {Marois}, \& {Barman}}]{lavigne2009}
{Lavigne}, J.-F., {Doyon}, R., {Lafreni{\`e}re}, D., {Marois}, C., \& {Barman},
  T. 2009, \apj, 704, 1098

\bibitem[{{Leya} {et~al.}(2008){Leya}, {Sch{\"o}nb{\"a}chler}, {Wiechert},
  {Kr{\"a}henb{\"u}hl}, \& {Halliday}}]{leya2008}
{Leya}, I., {Sch{\"o}nb{\"a}chler}, M., {Wiechert}, U., {Kr{\"a}henb{\"u}hl},
  U., \& {Halliday}, A.~N. 2008, Earth and Planetary Science Letters, 266, 233

\bibitem[{{Lincowski} {et~al.}(2019){Lincowski}, {Lustig-Yaeger}, \&
  {Meadows}}]{lincowski2019}
{Lincowski}, A.~P., {Lustig-Yaeger}, J., \& {Meadows}, V.~S. 2019, \aj, 158, 26

\bibitem[{{Linsky} {et~al.}(2006){Linsky}, {Draine}, {Moos}, {Jenkins}, {Wood},
  {Oliveira}, {Blair}, {Friedman}, {Gry}, {Knauth}, {Kruk}, {Lacour}, {Lehner},
  {Redfield}, {Shull}, {Sonneborn}, \& {Williger}}]{linsky2006}
{Linsky}, J.~L., {Draine}, B.~T., {Moos}, H.~W., {et~al.} 2006, \apj, 647, 1106

\bibitem[{Marconi {et~al.}(2020)Marconi, Abreu, Adibekyan, Aliverti, Prieto,
  Amado, Amate, Artigau, Augusto, Barros, Becerril, Benneke, Bergin, Berio,
  Bezawada, Boisse, Bonfils, Bouchy, Broeg, Cabral, Calvo-Ortega, Martins,
  Chazelas, Chiavassa, Christensen, Cirami, Coretti, Covino, Cresci, Cristiani,
  Parro, Cupani, D'Odorico, de~Castro~Leão, de~Medeiros, de~Souza,
  Marcantonio, Varano, Doyon, Drass, Figueira, Fragoso, Fynbo, Gallo, Genoni,
  Hernández, Gratton, Haehnelt, Hansen, Hlavacek-Larrondo, Hughes, Huke,
  Humphrey, Kjeldsen, Korn, Kouach, Landoni, Liske, Lovis, Lunney, Maiolino,
  Malo, Marquart, Martins, Maslowski, Mason, Micela, Molaro, Monnier, Monteiro,
  Mordasini, Morris, Mucciarelli, Murray, Niedzielski, Niemczura, Nisini,
  Nunes, Oliva, Origlia, Pallé, Pariani, Parr-Burman, Pasquini, Peñate, Pepe,
  Pietrzynski, Pinna, Piskunov, Pollo, Rasilla, Rebolo, II, Reiners, Riva,
  Romano, Rousseau, Sanna, Sarajlic, Shen, Sortino, Sosnowska, Sousa, Stempels,
  Strassmeier, Tenegi, Tozzi, Udry, Valenziano, Vanzi, Weber, Woche, Xompero,
  Zackrisson, \& Osorio}]{marconi2020}
Marconi, A., Abreu, M., Adibekyan, V., {et~al.} 2020, in Ground-based and
  Airborne Instrumentation for Astronomy VIII, ed. C.~J. Evans, J.~J. Bryant,
  \& K.~Motohara, Vol. 11447, International Society for Optics and Photonics
  (SPIE), 461 -- 472

\bibitem[{{McKemmish} {et~al.}(2019){McKemmish}, {Masseron}, {Hoeijmakers},
  {P{\'e}rez-Mesa}, {Grimm}, {Yurchenko}, \& {Tennyson}}]{mcKemmish2019}
{McKemmish}, L.~K., {Masseron}, T., {Hoeijmakers}, H.~J., {et~al.} 2019,
  \mnras, 488, 2836

\bibitem[{Meija {et~al.}(2016)Meija, Coplen, Berglund, Brand, Bièvre,
  Gröning, Holden, Irrgeher, Loss, Walczyk, \& Prohaska}]{meija2016}
Meija, J., Coplen, T.~B., Berglund, M., {et~al.} 2016, Pure and Applied
  Chemistry, 88, 293

\bibitem[{{Milam} {et~al.}(2005){Milam}, {Savage}, {Brewster}, {Ziurys}, \&
  {Wyckoff}}]{milam2005}
{Milam}, S.~N., {Savage}, C., {Brewster}, M.~A., {Ziurys}, L.~M., \& {Wyckoff},
  S. 2005, \apj, 634, 1126

\bibitem[{{Molli{\`e}re} \& {Snellen}(2019)}]{molliere2019a}
{Molli{\`e}re}, P. \& {Snellen}, I.~A.~G. 2019, \aap, 622, A139

\bibitem[{{Molli{\`e}re} {et~al.}(2015){Molli{\`e}re}, {van Boekel},
  {Dullemond}, {Henning}, \& {Mordasini}}]{molliere2015}
{Molli{\`e}re}, P., {van Boekel}, R., {Dullemond}, C., {Henning}, T., \&
  {Mordasini}, C. 2015, \apj, 813, 47

\bibitem[{{Molli{\`e}re} {et~al.}(2019){Molli{\`e}re}, {Wardenier}, {van
  Boekel}, {Henning}, {Molaverdikhani}, \& {Snellen}}]{molliere2019b}
{Molli{\`e}re}, P., {Wardenier}, J.~P., {van Boekel}, R., {et~al.} 2019, \aap,
  627, A67

\bibitem[{{Morley} {et~al.}(2019){Morley}, {Skemer}, {Miles}, {Line}, {Lopez},
  {Brogi}, {Freedman}, \& {Marley}}]{morley2019}
{Morley}, C.~V., {Skemer}, A.~J., {Miles}, B.~E., {et~al.} 2019, \apjl, 882,
  L29

\bibitem[{{Noll} {et~al.}(2012){Noll}, {Kausch}, {Barden}, {Jones}, {Szyszka},
  {Kimeswenger}, \& {Vinther}}]{noll2012}
{Noll}, S., {Kausch}, W., {Barden}, M., {et~al.} 2012, \aap, 543, A92

\bibitem[{{Nugroho} {et~al.}(2021){Nugroho}, {Kawahara}, {Gibson}, {de Mooij},
  {Hirano}, {Kotani}, {Kawashima}, {Masuda}, {Brogi}, {Birkby}, {Watson},
  {Tamura}, {Zwintz}, {Harakawa}, {Kudo}, {Kuzuhara}, {Hodapp}, {Ishizuka},
  {Jacobson}, {Konishi}, {Kurokawa}, {Nishikawa}, {Omiya}, {Serizawa}, {Ueda},
  \& {Vievard}}]{nugroho2021}
{Nugroho}, S.~K., {Kawahara}, H., {Gibson}, N.~P., {et~al.} 2021, \apjl, 910,
  L9

\bibitem[{{Nugroho} {et~al.}(2017){Nugroho}, {Kawahara}, {Masuda}, {Hirano},
  {Kotani}, \& {Tajitsu}}]{nugroho2017}
{Nugroho}, S.~K., {Kawahara}, H., {Masuda}, K., {et~al.} 2017, \aj, 154, 221

\bibitem[{{Pavlenko} {et~al.}(2020){Pavlenko}, {Yurchenko}, {McKemmish}, \&
  {Tennyson}}]{pavlenko2020}
{Pavlenko}, Y.~V., {Yurchenko}, S.~N., {McKemmish}, L.~K., \& {Tennyson}, J.
  2020, \aap, 642, A77

\bibitem[{{Quirrenbach} {et~al.}(2014){Quirrenbach}, {Amado}, {Caballero},
  {Mundt}, {Reiners}, {Ribas}, {Seifert}, {Abril}, {Aceituno},
  {Alonso-Floriano}, {Ammler-von Eiff}, {Antona Jim{\'e}nez},
  {Anwand-Heerwart}, {Azzaro}, {Bauer}, {Barrado}, {Becerril}, {B{\'e}jar},
  {Ben{\'\i}tez}, {Berdi{\~n}as}, {C{\'a}rdenas}, {Casal}, {Claret},
  {Colom{\'e}}, {Cort{\'e}s-Contreras}, {Czesla}, {Doellinger}, {Dreizler},
  {Feiz}, {Fern{\'a}ndez}, {Galad{\'\i}}, {G{\'a}lvez-Ortiz},
  {Garc{\'\i}a-Piquer}, {Garc{\'\i}a-Vargas}, {Garrido}, {Gesa}, {G{\'o}mez
  Galera}, {Gonz{\'a}lez {\'A}lvarez}, {Gonz{\'a}lez Hern{\'a}ndez},
  {Gr{\"o}zinger}, {Gu{\`a}rdia}, {Guenther}, {de Guindos},
  {Guti{\'e}rrez-Soto}, {Hagen}, {Hatzes}, {Hauschildt}, {Helmling}, {Henning},
  {Hermann}, {Hern{\'a}ndez Casta{\~n}o}, {Herrero}, {Hidalgo}, {Holgado},
  {Huber}, {Huber}, {Jeffers}, {Joergens}, {de Juan}, {Kehr}, {Klein},
  {K{\"u}rster}, {Lamert}, {Lalitha}, {Laun}, {Lemke}, {Lenzen}, {L{\'o}pez del
  Fresno}, {L{\'o}pez Mart{\'\i}}, {L{\'o}pez-Santiago}, {Mall}, {Mandel},
  {Mart{\'\i}n}, {Mart{\'\i}n-Ruiz}, {Mart{\'\i}nez-Rodr{\'\i}guez}, {Marvin},
  {Mathar}, {Mirabet}, {Montes}, {Morales Mu{\~n}oz}, {Moya}, {Naranjo},
  {Ofir}, {Oreiro}, {Pall{\'e}}, {Panduro}, {Passegger}, {P{\'e}rez-Calpena},
  {P{\'e}rez Medialdea}, {Perger}, {Pluto}, {Ram{\'o}n}, {Rebolo}, {Redondo},
  {Reffert}, {Reinhardt}, {Rhode}, {Rix}, {Rodler}, {Rodr{\'\i}guez},
  {Rodr{\'\i}guez-L{\'o}pez}, {Rodr{\'\i}guez-P{\'e}rez}, {Rohloff}, {Rosich},
  {S{\'a}nchez-Blanco}, {S{\'a}nchez Carrasco}, {Sanz-Forcada}, {Sarmiento},
  {Sch{\"a}fer}, {Schiller}, {Schmidt}, {Schmitt}, {Solano}, {Stahl}, {Storz},
  {St{\"u}rmer}, {Su{\'a}rez}, {Ulbrich}, {Veredas}, {Wagner}, {Winkler},
  {Zapatero Osorio}, {Zechmeister}, {Abell{\'a}n de Paco},
  {Anglada-Escud{\'e}}, {del Burgo}, {Klutsch}, {Lizon}, {L{\'o}pez-Morales},
  {Morales}, {Perryman}, {Tulloch}, \& {Xu}}]{quirrenbach2014}
{Quirrenbach}, A., {Amado}, P.~J., {Caballero}, J.~A., {et~al.} 2014, in
  Society of Photo-Optical Instrumentation Engineers (SPIE) Conference Series,
  Vol. 9147, Ground-based and Airborne Instrumentation for Astronomy V, ed.
  S.~K. {Ramsay}, I.~S. {McLean}, \& H.~{Takami}, 91471F

\bibitem[{{Rajpurohit} {et~al.}(2018){Rajpurohit}, {Allard}, {Rajpurohit},
  {Sharma}, {Teixeira}, {Mousis}, \& {Rajpurohit}}]{rajpurohit2018}
{Rajpurohit}, A.~S., {Allard}, F., {Rajpurohit}, S., {et~al.} 2018, \aap, 620,
  A180

\bibitem[{{Romano} {et~al.}(2017){Romano}, {Matteucci}, {Zhang},
  {Papadopoulos}, \& {Ivison}}]{romano2017}
{Romano}, D., {Matteucci}, F., {Zhang}, Z.~Y., {Papadopoulos}, P.~P., \&
  {Ivison}, R.~J. 2017, \mnras, 470, 401

\bibitem[{{Serindag} {et~al.}(2021){Serindag}, {Nugroho}, {Molli{\`e}re}, {de
  Mooij}, {Gibson}, \& {Snellen}}]{serindag2021}
{Serindag}, D.~B., {Nugroho}, S.~K., {Molli{\`e}re}, P., {et~al.} 2021, \aap,
  645, A90

\bibitem[{Spiegel {et~al.}(2009)Spiegel, Silverio, \& Burrows}]{spiegel2009}
Spiegel, D.~S., Silverio, K., \& Burrows, A. 2009, \apj, 699, 1487

\bibitem[{{Trinquier} {et~al.}(2009){Trinquier}, {Elliott}, {Ulfbeck}, {Coath},
  {Krot}, \& {Bizzarro}}]{trinquier2009}
{Trinquier}, A., {Elliott}, T., {Ulfbeck}, D., {et~al.} 2009, Science, 324, 374

\bibitem[{{Valenti} {et~al.}(1998){Valenti}, {Piskunov}, \&
  {Johns-Krull}}]{valenti1998}
{Valenti}, J.~A., {Piskunov}, N., \& {Johns-Krull}, C.~M. 1998, \apj, 498, 851

\bibitem[{{Walker}(1983)}]{walker1983}
{Walker}, A.~R. 1983, South African Astronomical Observatory Circular, 7, 106

\bibitem[{{Wood} {et~al.}(2004){Wood}, {Linsky}, {H{\'e}brard}, {Williger},
  {Moos}, \& {Blair}}]{wood2004}
{Wood}, B.~E., {Linsky}, J.~L., {H{\'e}brard}, G., {et~al.} 2004, \apj, 609,
  838

\bibitem[{{Wu} {et~al.}(2017){Wu}, {Sheehan}, {Males}, {Close}, {Morzinski},
  {Teske}, {Haug-Baltzell}, {Merchant}, \& {Lyons}}]{wu2017}
{Wu}, Y.-L., {Sheehan}, P.~D., {Males}, J.~R., {et~al.} 2017, \apj, 836, 223

\bibitem[{{Wyckoff} \& {Wehinger}(1972)}]{wyckoff1972}
{Wyckoff}, S. \& {Wehinger}, P. 1972, \apj, 178, 481

\bibitem[{{Zechmeister, M.} {et~al.}(2018){Zechmeister, M.}, {Reiners, A.},
  {Amado, P. J.}, {Azzaro, M.}, {Bauer, F. F.}, {B\'ejar, V. J. S.},
  {Caballero, J. A.}, {Guenther, E. W.}, {Hagen, H.-J.}, {Jeffers, S. V.},
  {Kaminski, A.}, {K\"urster, M.}, {Launhardt, R.}, {Montes, D.}, {Morales, J.
  C.}, {Quirrenbach, A.}, {Reffert, S.}, {Ribas, I.}, {Seifert, W.}, {Tal-Or,
  L.}, \& {Wolthoff, V.}}]{zechmeister2018}
{Zechmeister, M.}, {Reiners, A.}, {Amado, P. J.}, {et~al.} 2018, A\&A, 609, A12

\bibitem[{{Zhang} {et~al.}(2012){Zhang}, {Dauphas}, {Davis}, {Leya}, \&
  {Fedkin}}]{zhang2012}
{Zhang}, J., {Dauphas}, N., {Davis}, A.~M., {Leya}, I., \& {Fedkin}, A. 2012,
  Nature Geoscience, 5, 251

\bibitem[{{Zhang} {et~al.}(2021){Zhang}, {Snellen}, {Bohn}, {Molli{\`e}re},
  {Ginski}, {Hoeijmakers}, {Kenworthy}, {Mamajek}, {Meshkat}, {Reggiani}, \&
  {Snik}}]{zhang2021}
{Zhang}, Y., {Snellen}, I. A.~G., {Bohn}, A.~J., {et~al.} 2021, \nat, 595, 370

\end{thebibliography}

\begin{appendix}

\section{Supplementary Materials} \label{appendix}


\begin{figure*}
    \centering
        \includegraphics[width=17cm]{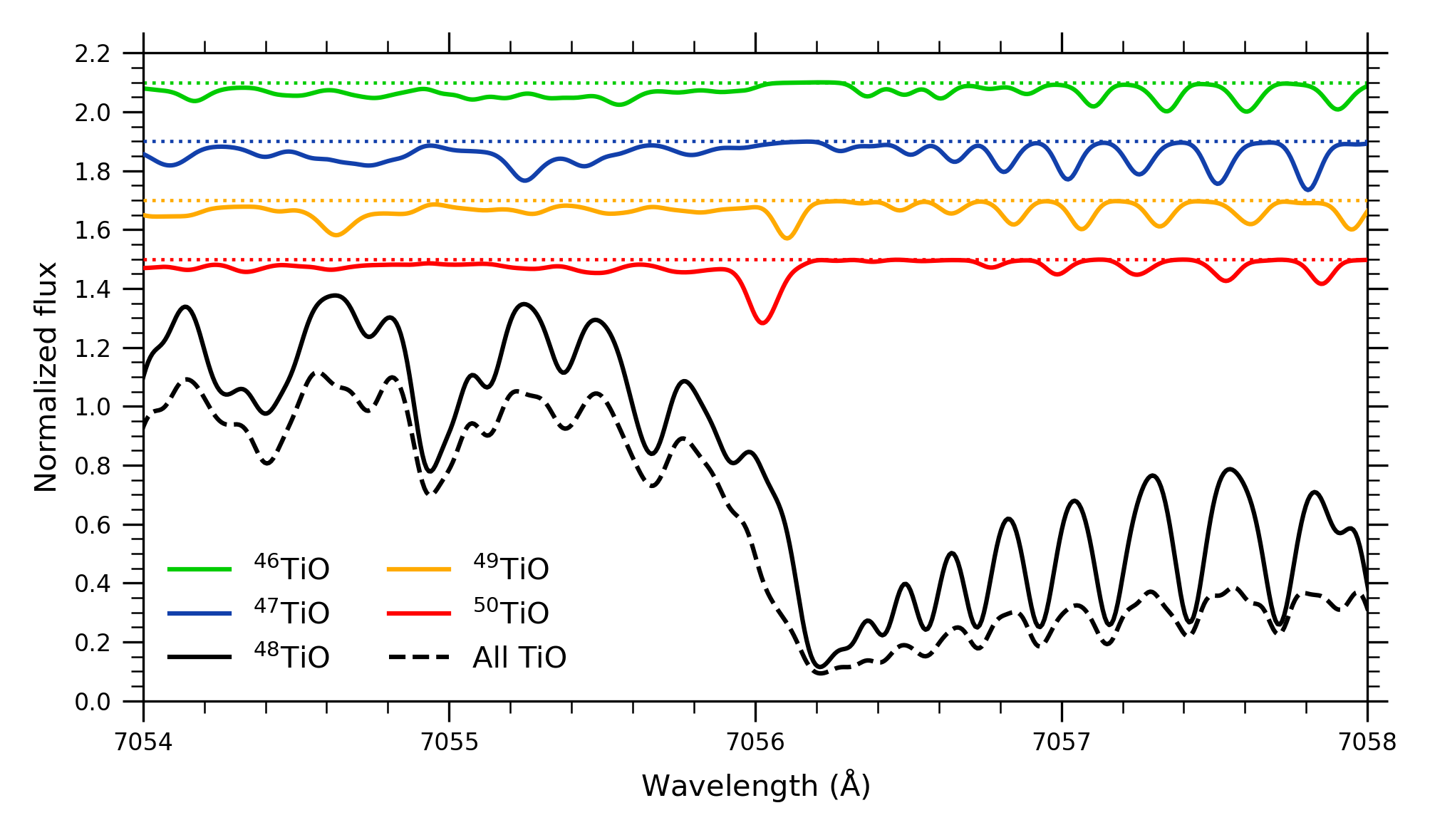}
        \caption{Flux contribution of each TiO isotopologue (solid lines) to the full TiO spectral model containing all isotopologues (dashed black line) around the strong band head at 7056 \AA{} (7054 \AA{} in air). Abundances are set at the retrieved values presented in Table \ref{table:method_parameters} for the narrow and unfiltered \gj{} spectrum. The flux contribution of a given minor isotopologue \isotope{TiO}{$i$} is calculated as the difference between the model of all TiO isotopologues and the model of all isotopologues except \isotope{TiO}{$i$}. Since \isotope{TiO}{48} comprises 63\% of TiO molecules in this setup, its flux contribution (solid black line) is calculated individually, without the presence of the minor isotopologues. Each contribution is normalized based on the full model's flux to ensure consistent scaling. The minor isotopologue contributions are vertically shifted to aid readability, with the offsets indicated by the dotted lines.}
        \label{fig:iso}
\end{figure*}


\begin{table*}
\caption{Mean parameter values and 1\sig{} errors from MCMC retrievals using different wavelength ranges and broadband filtering}
\label{table:method_parameters}
\centering
\begin{tabular}{l l l l l l l l}
Parameter & Value & Parameter & Value & Parameter & Value & Parameter & Value \\
\hline\hline
\multicolumn{8}{l}{Narrow/unfiltered spectrum} \\
\hline
$ \log_{10}{( \mathrm{VMR}_{46} )} $ & $ -8.60 \pm 0.01 $&  $ \textsuperscript{46}\mathrm{Ti} / \mathrm{Ti} $ & $ 0.104 \pm 0.003 $&  $ \textsuperscript{46}\mathrm{Ti} / \textsuperscript{48}\mathrm{Ti} $ & $ 0.165 \pm 0.006 $&  --        &        \\
$ \log_{10}{( \mathrm{VMR}_{47} )} $ & $ -8.56 \pm 0.01 $&  $ \textsuperscript{47}\mathrm{Ti} / \mathrm{Ti}  $ & $ 0.114^{+0.004}_{-0.003} $&  $ \textsuperscript{47}\mathrm{Ti} / \textsuperscript{48}\mathrm{Ti}  $ & $ 0.180^{+0.007}_{-0.006} $&  $ \textsuperscript{47}\mathrm{Ti} / \textsuperscript{46}\mathrm{Ti} $ & $ 1.09 \pm 0.05 $ \\
$ \log_{10}{( \mathrm{VMR}_{48} )} $ & $ -7.81 \pm 0.01 $&  $ \textsuperscript{48}\mathrm{Ti} / \mathrm{Ti}  $ & $ 0.630 \pm 0.006 $&  --        &       &  $ \textsuperscript{48}\mathrm{Ti} / \textsuperscript{46}\mathrm{Ti}  $ & $ 6.1 \pm 0.2 $ \\
$ \log_{10}{( \mathrm{VMR}_{49} )} $ & $ -8.67 \pm 0.02 $&  $ \textsuperscript{49}\mathrm{Ti} / \mathrm{Ti}  $ & $ 0.087 \pm 0.003 $&  $ \textsuperscript{49}\mathrm{Ti} / \textsuperscript{48}\mathrm{Ti}  $ & $ 0.138 \pm 0.006 $&  $ \textsuperscript{49}\mathrm{Ti} / \textsuperscript{46}\mathrm{Ti}  $ & $ 0.83 \pm 0.04 $ \\
$ \log_{10}{( \mathrm{VMR}_{50} )} $ & $ -8.80 \pm 0.02 $&  $ \textsuperscript{50}\mathrm{Ti} / \mathrm{Ti}  $ & $ 0.065 \pm 0.003 $&  $ \textsuperscript{50}\mathrm{Ti} / \textsuperscript{48}\mathrm{Ti}  $ & $ 0.103 \pm 0.005 $&  $ \textsuperscript{50}\mathrm{Ti} / \textsuperscript{46}\mathrm{Ti}  $ & $ 0.63^{+0.04}_{-0.03} $ \\
$\beta{}$ & $ 4.94^{+0.07}_{-0.06} $&  --        &       &  --        &       &  --        &        \\
\hline
\multicolumn{8}{l}{Narrow/filtered spectrum} \\
\hline
$ \log_{10}{( \mathrm{VMR}_{46} )} $ & $ -8.47 \pm 0.02 $&  $ \textsuperscript{46}\mathrm{Ti} / \mathrm{Ti} $ & $ 0.109 \pm 0.003 $&  $ \textsuperscript{46}\mathrm{Ti} / \textsuperscript{48}\mathrm{Ti} $ & $ 0.172 \pm 0.006 $&  --        &        \\
$ \log_{10}{( \mathrm{VMR}_{47} )} $ & $ -8.48 \pm 0.02 $&  $ \textsuperscript{47}\mathrm{Ti} / \mathrm{Ti}  $ & $ 0.107 \pm 0.003 $&  $ \textsuperscript{47}\mathrm{Ti} / \textsuperscript{48}\mathrm{Ti}  $ & $ 0.169 \pm 0.006 $&  $ \textsuperscript{47}\mathrm{Ti} / \textsuperscript{46}\mathrm{Ti} $ & $ 0.98 \pm 0.03 $ \\
$ \log_{10}{( \mathrm{VMR}_{48} )} $ & $ -7.71^{+0.02}_{-0.03} $&  $ \textsuperscript{48}\mathrm{Ti} / \mathrm{Ti}  $ & $ 0.635 \pm 0.006 $&  --        &       &  $ \textsuperscript{48}\mathrm{Ti} / \textsuperscript{46}\mathrm{Ti}  $ & $ 5.8 \pm 0.2 $ \\
$ \log_{10}{( \mathrm{VMR}_{49} )} $ & $ -8.60 \pm 0.02 $&  $ \textsuperscript{49}\mathrm{Ti} / \mathrm{Ti}  $ & $ 0.081 \pm 0.003 $&  $ \textsuperscript{49}\mathrm{Ti} / \textsuperscript{48}\mathrm{Ti}  $ & $ 0.128^{+0.005}_{-0.004} $&  $ \textsuperscript{49}\mathrm{Ti} / \textsuperscript{46}\mathrm{Ti}  $ & $ 0.75 \pm 0.03 $ \\
$ \log_{10}{( \mathrm{VMR}_{50} )} $ & $ -8.68^{+0.02}_{-0.03} $&  $ \textsuperscript{50}\mathrm{Ti} / \mathrm{Ti}  $ & $ 0.068^{+0.002}_{-0.003} $&  $ \textsuperscript{50}\mathrm{Ti} / \textsuperscript{48}\mathrm{Ti}  $ & $ 0.107^{+0.004}_{-0.005} $&  $ \textsuperscript{50}\mathrm{Ti} / \textsuperscript{46}\mathrm{Ti}  $ & $ 0.62 \pm 0.03 $ \\
$\beta{}$ & $ 3.87 \pm 0.05 $&  --        &       &  --        &       &  --        &        \\
\hline
\multicolumn{8}{l}{Wide/unfiltered spectrum} \\
\hline
$ \log_{10}{( \mathrm{VMR}_{46} )} $ & $ -8.657 \pm 0.009 $&  $ \textsuperscript{46}\mathrm{Ti} / \mathrm{Ti} $ & $ 0.106 \pm 0.002 $&  $ \textsuperscript{46}\mathrm{Ti} / \textsuperscript{48}\mathrm{Ti} $ & $ 0.168 \pm 0.004 $&  --        &        \\
$ \log_{10}{( \mathrm{VMR}_{47} )} $ & $ -8.635 \pm 0.009 $&  $ \textsuperscript{47}\mathrm{Ti} / \mathrm{Ti}  $ & $ 0.112 \pm 0.002 $&  $ \textsuperscript{47}\mathrm{Ti} / \textsuperscript{48}\mathrm{Ti}  $ & $ 0.177 \pm 0.004 $&  $ \textsuperscript{47}\mathrm{Ti} / \textsuperscript{46}\mathrm{Ti} $ & $ 1.05 \pm 0.03 $ \\
$ \log_{10}{( \mathrm{VMR}_{48} )} $ & $ -7.883 \pm 0.004 $&  $ \textsuperscript{48}\mathrm{Ti} / \mathrm{Ti}  $ & $ 0.631 \pm 0.003 $&  --        &       &  $ \textsuperscript{48}\mathrm{Ti} / \textsuperscript{46}\mathrm{Ti}  $ & $ 5.9 \pm 0.1 $ \\
$ \log_{10}{( \mathrm{VMR}_{49} )} $ & $ -8.77 \pm 0.01 $&  $ \textsuperscript{49}\mathrm{Ti} / \mathrm{Ti}  $ & $ 0.081 \pm 0.002 $&  $ \textsuperscript{49}\mathrm{Ti} / \textsuperscript{48}\mathrm{Ti}  $ & $ 0.129 \pm 0.004 $&  $ \textsuperscript{49}\mathrm{Ti} / \textsuperscript{46}\mathrm{Ti}  $ & $ 0.77 \pm 0.03 $ \\
$ \log_{10}{( \mathrm{VMR}_{50} )} $ & $ -8.84 \pm 0.01 $&  $ \textsuperscript{50}\mathrm{Ti} / \mathrm{Ti}  $ & $ 0.069 \pm 0.002 $&  $ \textsuperscript{50}\mathrm{Ti} / \textsuperscript{48}\mathrm{Ti}  $ & $ 0.110^{+0.004}_{-0.003} $&  $ \textsuperscript{50}\mathrm{Ti} / \textsuperscript{46}\mathrm{Ti}  $ & $ 0.66 \pm 0.03 $ \\
$\beta{}$ & $ 7.18 \pm 0.03 $&  --        &       &  --        &       &  --        &        \\
\hline
\multicolumn{8}{l}{Wide/filtered spectrum} \\
\hline
$ \log_{10}{( \mathrm{VMR}_{46} )} $ & $ -8.623^{+0.01}_{-0.010} $&  $ \textsuperscript{46}\mathrm{Ti} / \mathrm{Ti} $ & $ 0.101 \pm 0.002 $&  $ \textsuperscript{46}\mathrm{Ti} / \textsuperscript{48}\mathrm{Ti} $ & $ 0.152 \pm 0.003 $&  --        &        \\
$ \log_{10}{( \mathrm{VMR}_{47} )} $ & $ -8.617^{+0.009}_{-0.010} $&  $ \textsuperscript{47}\mathrm{Ti} / \mathrm{Ti}  $ & $ 0.102 \pm 0.002 $&  $ \textsuperscript{47}\mathrm{Ti} / \textsuperscript{48}\mathrm{Ti}  $ & $ 0.154 \pm 0.003 $&  $ \textsuperscript{47}\mathrm{Ti} / \textsuperscript{46}\mathrm{Ti} $ & $ 1.01 \pm 0.03 $ \\
$ \log_{10}{( \mathrm{VMR}_{48} )} $ & $ -7.805^{+0.007}_{-0.006} $&  $ \textsuperscript{48}\mathrm{Ti} / \mathrm{Ti}  $ & $ 0.664 \pm 0.003 $&  --        &       &  $ \textsuperscript{48}\mathrm{Ti} / \textsuperscript{46}\mathrm{Ti}  $ & $ 6.6 \pm 0.1 $ \\
$ \log_{10}{( \mathrm{VMR}_{49} )} $ & $ -8.76 \pm 0.01 $&  $ \textsuperscript{49}\mathrm{Ti} / \mathrm{Ti}  $ & $ 0.074 \pm 0.002 $&  $ \textsuperscript{49}\mathrm{Ti} / \textsuperscript{48}\mathrm{Ti}  $ & $ 0.111 \pm 0.003 $&  $ \textsuperscript{49}\mathrm{Ti} / \textsuperscript{46}\mathrm{Ti}  $ & $ 0.73 \pm 0.02 $ \\
$ \log_{10}{( \mathrm{VMR}_{50} )} $ & $ -8.86 \pm 0.01 $&  $ \textsuperscript{50}\mathrm{Ti} / \mathrm{Ti}  $ & $ 0.059 \pm 0.001 $&  $ \textsuperscript{50}\mathrm{Ti} / \textsuperscript{48}\mathrm{Ti}  $ & $ 0.088 \pm 0.002 $&  $ \textsuperscript{50}\mathrm{Ti} / \textsuperscript{46}\mathrm{Ti}  $ & $ 0.58 \pm 0.02 $ \\
$\beta{}$ & $ 4.81^{+0.02}_{-0.03} $&  --        &       &  --        &       &  --        &        \\
\end{tabular}
\end{table*}


\begin{table*}
\caption{Formal deviation between parameters retrieved using different wavelength ranges and broadband filtering}
\label{table:method_agreement}
\centering
\begin{tabular}{l l l l l l l l}
Parameter & Deviation ($\sigma{}$) & Parameter & Deviation ($\sigma{}$) & Parameter & Deviation ($\sigma{}$) & Parameter & Deviation ($\sigma{}$) \\
\hline\hline
\multicolumn{8}{l}{Narrow/unfiltered vs. narrow/filtered} \\
\hline
$ \log_{10}{( \mathrm{VMR}_{46} )} $ & 4.75 &  $ \textsuperscript{46}\mathrm{Ti} / \mathrm{Ti} $ & 1.11 &  $ \textsuperscript{46}\mathrm{Ti} / \textsuperscript{48}\mathrm{Ti} $ & 0.79 &  --        &        \\
$ \log_{10}{( \mathrm{VMR}_{47} )} $ & 2.95 &  $ \textsuperscript{47}\mathrm{Ti} / \mathrm{Ti}  $ & 1.44 &  $ \textsuperscript{47}\mathrm{Ti} / \textsuperscript{48}\mathrm{Ti}  $ & 1.34 &  $ \textsuperscript{47}\mathrm{Ti} / \textsuperscript{46}\mathrm{Ti} $ & 1.88  \\
$ \log_{10}{( \mathrm{VMR}_{48} )} $ & 3.86 &  $ \textsuperscript{48}\mathrm{Ti} / \mathrm{Ti}  $ & 0.58 &  --        &       &  $ \textsuperscript{48}\mathrm{Ti} / \textsuperscript{46}\mathrm{Ti}  $ & 0.79  \\
$ \log_{10}{( \mathrm{VMR}_{49} )} $ & 2.64 &  $ \textsuperscript{49}\mathrm{Ti} / \mathrm{Ti}  $ & 1.31 &  $ \textsuperscript{49}\mathrm{Ti} / \textsuperscript{48}\mathrm{Ti}  $ & 1.28 &  $ \textsuperscript{49}\mathrm{Ti} / \textsuperscript{46}\mathrm{Ti}  $ & 1.66  \\
$ \log_{10}{( \mathrm{VMR}_{50} )} $ & 3.71 &  $ \textsuperscript{50}\mathrm{Ti} / \mathrm{Ti}  $ & 0.64 &  $ \textsuperscript{50}\mathrm{Ti} / \textsuperscript{48}\mathrm{Ti}  $ & 0.46 &  $ \textsuperscript{50}\mathrm{Ti} / \textsuperscript{46}\mathrm{Ti}  $ & 0.14  \\
\hline
\multicolumn{8}{l}{Wide/unfiltered vs. wide/filtered} \\
\hline
$ \log_{10}{( \mathrm{VMR}_{46} )} $ & 2.59 &  $ \textsuperscript{46}\mathrm{Ti} / \mathrm{Ti} $ & 1.81 &  $ \textsuperscript{46}\mathrm{Ti} / \textsuperscript{48}\mathrm{Ti} $ & 3.32 &  --        &        \\
$ \log_{10}{( \mathrm{VMR}_{47} )} $ & 1.39 &  $ \textsuperscript{47}\mathrm{Ti} / \mathrm{Ti}  $ & 3.25 &  $ \textsuperscript{47}\mathrm{Ti} / \textsuperscript{48}\mathrm{Ti}  $ & 4.63 &  $ \textsuperscript{47}\mathrm{Ti} / \textsuperscript{46}\mathrm{Ti} $ & 0.93  \\
$ \log_{10}{( \mathrm{VMR}_{48} )} $ & 10.31&  $ \textsuperscript{48}\mathrm{Ti} / \mathrm{Ti}  $ & 8.11 &  --        &       &  $ \textsuperscript{48}\mathrm{Ti} / \textsuperscript{46}\mathrm{Ti}  $ & 3.35  \\
$ \log_{10}{( \mathrm{VMR}_{49} )} $ & 0.75 &  $ \textsuperscript{49}\mathrm{Ti} / \mathrm{Ti}  $ & 2.96 &  $ \textsuperscript{49}\mathrm{Ti} / \textsuperscript{48}\mathrm{Ti}  $ & 4.01 &  $ \textsuperscript{49}\mathrm{Ti} / \textsuperscript{46}\mathrm{Ti}  $ & 1.14  \\
$ \log_{10}{( \mathrm{VMR}_{50} )} $ & 1.05 &  $ \textsuperscript{50}\mathrm{Ti} / \mathrm{Ti}  $ & 4.38 &  $ \textsuperscript{50}\mathrm{Ti} / \textsuperscript{48}\mathrm{Ti}  $ & 5.32 &  $ \textsuperscript{50}\mathrm{Ti} / \textsuperscript{46}\mathrm{Ti}  $ & 2.38  \\
\hline
\multicolumn{8}{l}{Narrow/unfiltered vs. wide/unfiltered} \\
\hline
$ \log_{10}{( \mathrm{VMR}_{46} )} $ & 3.83 &  $ \textsuperscript{46}\mathrm{Ti} / \mathrm{Ti} $ & 0.54 &  $ \textsuperscript{46}\mathrm{Ti} / \textsuperscript{48}\mathrm{Ti} $ & 0.43 &  --        &        \\
$ \log_{10}{( \mathrm{VMR}_{47} )} $ & 4.87 &  $ \textsuperscript{47}\mathrm{Ti} / \mathrm{Ti}  $ & 0.49 &  $ \textsuperscript{47}\mathrm{Ti} / \textsuperscript{48}\mathrm{Ti}  $ & 0.46 &  $ \textsuperscript{47}\mathrm{Ti} / \textsuperscript{46}\mathrm{Ti} $ & 0.70  \\
$ \log_{10}{( \mathrm{VMR}_{48} )} $ & 5.89 &  $ \textsuperscript{48}\mathrm{Ti} / \mathrm{Ti}  $ & 0.18 &  --        &       &  $ \textsuperscript{48}\mathrm{Ti} / \textsuperscript{46}\mathrm{Ti}  $ & 0.43  \\
$ \log_{10}{( \mathrm{VMR}_{49} )} $ & 5.10 &  $ \textsuperscript{49}\mathrm{Ti} / \mathrm{Ti}  $ & 1.46 &  $ \textsuperscript{49}\mathrm{Ti} / \textsuperscript{48}\mathrm{Ti}  $ & 1.32 &  $ \textsuperscript{49}\mathrm{Ti} / \textsuperscript{46}\mathrm{Ti}  $ & 1.38  \\
$ \log_{10}{( \mathrm{VMR}_{50} )} $ & 1.89 &  $ \textsuperscript{50}\mathrm{Ti} / \mathrm{Ti}  $ & 1.21 &  $ \textsuperscript{50}\mathrm{Ti} / \textsuperscript{48}\mathrm{Ti}  $ & 1.09 &  $ \textsuperscript{50}\mathrm{Ti} / \textsuperscript{46}\mathrm{Ti}  $ & 0.64  \\
\hline
\multicolumn{8}{l}{Narrow/filtered vs. wide/filtered} \\
\hline
$ \log_{10}{( \mathrm{VMR}_{46} )} $ & 6.20 &  $ \textsuperscript{46}\mathrm{Ti} / \mathrm{Ti} $ & 2.24 &  $ \textsuperscript{46}\mathrm{Ti} / \textsuperscript{48}\mathrm{Ti} $ & 3.03 &  --        &        \\
$ \log_{10}{( \mathrm{VMR}_{47} )} $ & 5.70 &  $ \textsuperscript{47}\mathrm{Ti} / \mathrm{Ti}  $ & 1.34 &  $ \textsuperscript{47}\mathrm{Ti} / \textsuperscript{48}\mathrm{Ti}  $ & 2.28 &  $ \textsuperscript{47}\mathrm{Ti} / \textsuperscript{46}\mathrm{Ti} $ & 0.68  \\
$ \log_{10}{( \mathrm{VMR}_{48} )} $ & 3.79 &  $ \textsuperscript{48}\mathrm{Ti} / \mathrm{Ti}  $ & 4.79 &  --        &       &  $ \textsuperscript{48}\mathrm{Ti} / \textsuperscript{46}\mathrm{Ti}  $ & 3.15  \\
$ \log_{10}{( \mathrm{VMR}_{49} )} $ & 6.07 &  $ \textsuperscript{49}\mathrm{Ti} / \mathrm{Ti}  $ & 2.62 &  $ \textsuperscript{49}\mathrm{Ti} / \textsuperscript{48}\mathrm{Ti}  $ & 3.40 &  $ \textsuperscript{49}\mathrm{Ti} / \textsuperscript{46}\mathrm{Ti}  $ & 0.49  \\
$ \log_{10}{( \mathrm{VMR}_{50} )} $ & 6.34 &  $ \textsuperscript{50}\mathrm{Ti} / \mathrm{Ti}  $ & 3.06 &  $ \textsuperscript{50}\mathrm{Ti} / \textsuperscript{48}\mathrm{Ti}  $ & 3.61 &  $ \textsuperscript{50}\mathrm{Ti} / \textsuperscript{46}\mathrm{Ti}  $ & 1.21  \\
\hline
\multicolumn{8}{l}{Narrow/unfiltered vs. wide/filtered} \\
\hline
$ \log_{10}{( \mathrm{VMR}_{46} )} $ & 1.63 &  $ \textsuperscript{46}\mathrm{Ti} / \mathrm{Ti} $ & 0.82 &  $ \textsuperscript{46}\mathrm{Ti} / \textsuperscript{48}\mathrm{Ti} $ & 1.98 &  --        &        \\
$ \log_{10}{( \mathrm{VMR}_{47} )} $ & 3.72 &  $ \textsuperscript{47}\mathrm{Ti} / \mathrm{Ti}  $ & 3.00 &  $ \textsuperscript{47}\mathrm{Ti} / \textsuperscript{48}\mathrm{Ti}  $ & 3.74 &  $ \textsuperscript{47}\mathrm{Ti} / \textsuperscript{46}\mathrm{Ti} $ & 1.47  \\
$ \log_{10}{( \mathrm{VMR}_{48} )} $ & 0.62 &  $ \textsuperscript{48}\mathrm{Ti} / \mathrm{Ti}  $ & 5.29 &  --        &       &  $ \textsuperscript{48}\mathrm{Ti} / \textsuperscript{46}\mathrm{Ti}  $ & 2.02  \\
$ \log_{10}{( \mathrm{VMR}_{49} )} $ & 4.51 &  $ \textsuperscript{49}\mathrm{Ti} / \mathrm{Ti}  $ & 3.90 &  $ \textsuperscript{49}\mathrm{Ti} / \textsuperscript{48}\mathrm{Ti}  $ & 4.40 &  $ \textsuperscript{49}\mathrm{Ti} / \textsuperscript{46}\mathrm{Ti}  $ & 2.33  \\
$ \log_{10}{( \mathrm{VMR}_{50} )} $ & 2.75 &  $ \textsuperscript{50}\mathrm{Ti} / \mathrm{Ti}  $ & 2.15 &  $ \textsuperscript{50}\mathrm{Ti} / \textsuperscript{48}\mathrm{Ti}  $ & 2.91 &  $ \textsuperscript{50}\mathrm{Ti} / \textsuperscript{46}\mathrm{Ti}  $ & 1.23  \\
\hline
\multicolumn{8}{l}{Narrow/filtered vs. wide/unfiltered} \\
\hline
$ \log_{10}{( \mathrm{VMR}_{46} )} $ & 7.74 &  $ \textsuperscript{46}\mathrm{Ti} / \mathrm{Ti} $ & 0.76 &  $ \textsuperscript{46}\mathrm{Ti} / \textsuperscript{48}\mathrm{Ti} $ & 0.51 &  --        &        \\
$ \log_{10}{( \mathrm{VMR}_{47} )} $ & 6.45 &  $ \textsuperscript{47}\mathrm{Ti} / \mathrm{Ti}  $ & 1.19 &  $ \textsuperscript{47}\mathrm{Ti} / \textsuperscript{48}\mathrm{Ti}  $ & 1.15 &  $ \textsuperscript{47}\mathrm{Ti} / \textsuperscript{46}\mathrm{Ti} $ & 1.45  \\
$ \log_{10}{( \mathrm{VMR}_{48} )} $ & 6.89 &  $ \textsuperscript{48}\mathrm{Ti} / \mathrm{Ti}  $ & 0.55 &  --        &       &  $ \textsuperscript{48}\mathrm{Ti} / \textsuperscript{46}\mathrm{Ti}  $ & 0.50  \\
$ \log_{10}{( \mathrm{VMR}_{49} )} $ & 6.49 &  $ \textsuperscript{49}\mathrm{Ti} / \mathrm{Ti}  $ & 0.00 &  $ \textsuperscript{49}\mathrm{Ti} / \textsuperscript{48}\mathrm{Ti}  $ & 0.13 &  $ \textsuperscript{49}\mathrm{Ti} / \textsuperscript{46}\mathrm{Ti}  $ & 0.45  \\
$ \log_{10}{( \mathrm{VMR}_{50} )} $ & 5.66 &  $ \textsuperscript{50}\mathrm{Ti} / \mathrm{Ti}  $ & 0.56 &  $ \textsuperscript{50}\mathrm{Ti} / \textsuperscript{48}\mathrm{Ti}  $ & 0.62 &  $ \textsuperscript{50}\mathrm{Ti} / \textsuperscript{46}\mathrm{Ti}  $ & 0.88  \\
\end{tabular}
\end{table*}

\end{appendix}

\end{document}